\documentclass[12pt]{article}
\usepackage{amsmath,enumerate,amsfonts,color,amssymb, amsthm}
\usepackage{graphicx}
\usepackage{dcolumn}
\usepackage{bm}

\usepackage{tikz}
\usepackage{subfigure}

\DeclareFontFamily{OT1}{rsfs}{}
\DeclareFontShape{OT1}{rsfs}{m}{n}{ <-7> rsfs5 <7-10> rsfs7 <10-> rsfs10}{}
\DeclareMathAlphabet{\mycal}{OT1}{rsfs}{m}{n}

\newcommand{\bfmath}[1]{{\mathchoice
    {\mbox{\boldmath$\displaystyle#1$}}
    {\mbox{\boldmath$\textstyle#1$}}
    {\mbox{\boldmath$\scriptstyle#1$}}
    {\mbox{\boldmath$\scriptscriptstyle#1$}}}}

\def\tens#1{\ensuremath{\mathsf{#1}}}
   
\newtheorem{theorem} {\sc  Theorem\rm} [section]

\newtheorem{conjecture}[theorem]{\sc  Conjecture\rm}

\newtheorem{remark}[theorem]{\sc  Remark\rm}

\begin{document}

\title{On the Disclination Lines of Nematic Liquid Crystals}
\author{Yucheng Hu\thanks{Zhou Pei-yuan Center for Applied Mathematics, Tsinghua University, Beijing, China. Email: huyc@tsinghua.edu.cn}, Yang Qu$^{\dag}$ and Pingwen Zhang\thanks{Laboratory of Mathematics and Applied Mathematics and School of Mathematical Sciences, Peking University, Beijing, China. Email: quyang@pku.edu.cn, pzhang@pku.edu.cn}}
\date{}

\maketitle

\begin{abstract}
Defects in liquid crystals are of great practical importance and theoretical interest. Despite tremendous efforts, predicting the location and transition of defects under various topological constraint and external field remains to be a challenge. We investigate defect patterns of nematic liquid crystals confined in three-dimensional spherical droplet and two-dimensional disk under different boundary conditions, within the Landau-de Gennes model. We implement a spectral method that numerically solves the Landau-de Gennes model with high accuracy, which allows us to study the detailed static structure of defects. We observe five types of defect structures. Among them the 1/2-disclination lines are the most stable structure at low temperature. Inspired by numerical results, we obtain the profile of disclination lines analytically. Moreover, the connection and difference between defect patterns under the Landau-de Gennes model and the Oseen-Frank model are discussed. Finally, four conjectures are made to summarize some important characteristics of defects in the Landau-de Gennes theory. This work is a continuing effort to deepen our understanding on defect patterns in nematic liquid crystals.
\end{abstract}

\section{Introduction}
Nematic liquid crystals (LCs)  are composed of rigid rod-like molecules. When subject to topological constraint, discontinuity in the alignment direction of LCs can form, which is known as \emph{defects}. Defects are commonly found to exist as isolated point or disclination line in experiments~\cite{de1993physics}. When conditions such as temperature and boundary constraint vary, the location and topology of defects may change drastically~\cite{gupta2009size}. Predicting defect pattern is the key to design self-assembly biomolecule and colloidal suspensions, and is thus of particular practical interest but remains to be a difficult problem~\cite{miller2013design, lopez2011drops, erdmann1990configuration, lin2001static}.

Three commonly used continuum theories to describe nematic LCs at equilibrium are the Oseen-Frank model, Ericksen's model and the Landau-de Gennes model~\cite{lin2001static}. In the Oseen-Frank model the state of nematic LCs is described by a unit-vector field, $\bfmath{n} \in W^{1, 2}(\Omega; \mathbb{S}^{2})$, where $\Omega \in \mathbb{R}^d, d=2, 3$ is the region occupied by the LCs material.
In its simplest form, the Oseen-Frank free-energy functional can be written as
\begin{equation*}
 F_{OF}[\bfmath{n}] = \int_{\Omega} |\nabla \bfmath{n}|^2 d\bfmath{x}.
\end{equation*}
The vector filed $\bfmath{n}$ that minimizes $F_{OF}$ is a $\mathbb{S}^2$-valued harmonic map~\cite{lin2001static, virga1995variational}. 

There are two deficiencies in the Oseen-Frank model in describing nematic LCs. First, $\bfmath{n}$ and $-\bfmath{n}$ are treated as discontinuity while physically they are equivalent. As a result, the \emph{head-to-tail} symmetry is not preserved~\cite{ball2011orientability}. Secondly, the model can only predict point defects but not the more complex disclination lines observed in experiments~\cite{lavrentovich2003defects}.

The Ericksen's model can admit solutions that contain disclination lines~\cite{ericksen1991liquid, lin2001static}. In this model the state of LCs is described by $(s, \bfmath{n}) \in  W^{1, 2}(\Omega; \mathbb{R}\otimes\mathbb{S}^{2})$. Compared with the Oseen-Frank model, it contains an extra order parameter $s \in \mathbb{R}$ which measures the degree of orientational order along $\bfmath{n}$. The free-energy functional is given by
\begin{equation*}
 F_{E}[s, \bfmath{n}] = \int_{\Omega} s^2 |\nabla \bfmath{n}|^2 + k|\nabla s^2| + \omega_0 (s) d\bfmath{x},
\end{equation*}
where $\omega_0$ is a bulk energy term and $k$ is a constant. Singularity of $\bfmath{n}$ in $\mathbb{S}^{2}$ in the Oseen-Frank model at the defect can be removed by allowing $s=0$ in $\mathbb{R}\otimes\mathbb{S}^{2}$ in the Ericksen's model. In this sense the Ericksen's model can be considered as a regularization of the Oseen-Frank model.

In the physically more realistic Landau-de Gennes (LdG) model the state of LCs is described by a matrix-valued tensor field, $\tens{Q} \in W^{1, 2}(\Omega; \mycal{S}_0)$.  The set $\mycal{S}_0 := \{ \tens{Q} \in \mathbb{R}^{3\times 3}: \tens{Q} = \tens{Q}^T, \mathrm{tr}(\tens{Q}) = 0\}$ contains all the three-by-three symmetric traceless matrix. A tensor $\tens{Q} \in \mycal{S}_0$ has five degree-of-freedom and can be written as
\begin{equation}
 \tens{Q} = s\left( \bfmath{n}\bfmath{n} - \frac{\tens{I}}{3}\right) + r \left( \bfmath{m}\bfmath{m} - \frac{\tens{I}}{3} \right),  \ \ \ s, r \in \mathbb{R}, 
 \ \ \ \bfmath{n}, \bfmath{m} \in \mathbb{S}^2,
 \label{eq_Q}
\end{equation}
where $\tens{I}$ is three-by-three identity matrix. When $s = r = 0$, $\tens{Q} = 0$ and is called \emph{isotropic}. When $s\ne 0$ and $r = 0$, $\tens{Q} = s \left( \bfmath{n} \bfmath{n} - \frac{\tens{I}}{3}\right)$ is called \emph{uniaxial}. It corresponds to the physical configuration that the orientation of the LC molecules are rotational symmetrical with respect to $\bfmath{n}$. A uniaxial $\tens{Q}$ has two identical eigenvalues. 
The set of uniaxial and isotropic $\tens{Q}$,
\begin{equation}
 \mycal{U} := \{ \tens{Q} = s \left( \bfmath{n} \bfmath{n} - \frac{\tens{I}}{3}\right): s \in \mathbb{R}, \bfmath{n} \in \mathbb{S}^2\},
 \label{eq_uset}
\end{equation}
is homotopically equivalent to $\mathbb{R}\otimes\mathbb{S}^{2}$ for the order parameters $(s, \bfmath{n})$ in the Ericksen's model. In addition, for fixed $s = s^* \ne 0$ in $\Omega$, the order parameter reduce to $\bfmath{n}$ in the Oseen-Frank theory. When the three eigenvalues of $\tens{Q}$ are different, both $r$ and $s$ in Eq.~\eqref{eq_Q} are non-zero and $\tens{Q}$ is referred as \emph{biaxial}. As we will see later, biaxiality, which is absent in the Ericksen's and Oseen-Frank models, is a key ingredient in the local profile of defects in the LdG model.

The free-energy functional of the LdG model can be written as
\begin{equation*}
F[\tens{Q}] = \int_{\Omega}f_{b}(\tens{Q}) +  f_e (\tens{Q})\mathrm{d}V.
\end{equation*}
Here the bulk energy density is
\begin{equation}
f_b(\tens{Q}) = \frac{A}{2}\mathrm{tr}(\tens{Q}^2) - 
\frac{B}{3}\mathrm{tr}(\tens{Q}^3) + 
\frac{C}{4}\mathrm{tr}(\tens{Q}^2)^2,
\label{eq_fb}
\end{equation}
and the elastic energy density is
\begin{equation*}
f_e(\tens{Q}) =  \frac{L_1}{2} \tens{Q}_{ij,k} \tens{Q}_{ij,k} + \frac{L_2}{2} \tens{Q}_{ij,j} \tens{Q}_{ik,k} + \frac{L_3}{2} \tens{Q}_{ij,k} \tens{Q}_{ik,j}.
\end{equation*}
$A, B, C$ are temperature and material dependent constants and $L_1, L_2, L_3$ are elastic constants. Summation over repeated indices is implied and the comma indicates spatial derivative. For simplicity we restrict ourselves to $L_2 = L_3 = 0$ and consider the domain $\Omega$ as a 3-ball of radius $R$ ($\Omega = B_R$), or a 2-disk of radius $R$ ($\Omega = D_R$). We nondimensionalize the model by defining the characteristic length $\xi_0 = \sqrt{\frac{27CL_1}{B^2}}$, effective temperature $t = \frac{27AC}{B^2}$ and elastic constant $\varepsilon = \frac{\xi_0}{R}$, and rescaling the variables by $\tilde{\bfmath{x}} = \frac{\bfmath{x}}{R}, \bfmath{x} \in \Omega$, $\tilde{\tens{Q}} = \sqrt{\frac{27C^2}{2B^2}}\tens{Q}$,
$\tilde{F} = \varepsilon^d\sqrt{\frac{27C^3}{4 B^2L^3_1}} F$.
After dropping the tildes, we obtain
\begin{equation}
F[\tens{Q}] =\int_{\Omega} \frac{t}{2}\mathrm{tr}(\tens{Q}^2) - \sqrt{6}\mathrm{tr}(\tens{Q}^3) + \frac{1}{2}\mathrm{tr}(\tens{Q}^2)^2 + \frac{\varepsilon^2}{2}\tens{Q}_{ij,k}\tens{Q}_{ij,k} \mathrm{d}\bfmath{x}.
\label{eq_totalenergy}
\end{equation}
The integration is taken over the rescaled computational domain --- the unit ball ($\Omega = B_{R=1}$) or unit disk ($\Omega = D_{R=1}$). 

\begin{remark}
In a related work~\cite{mkaddem2000fine}, the length is rescaled by the characteristic length $\xi_0$ instead of $R$ as we did here. When $R$ increases in their case, the radius of the computation domain $\Omega$ also increases while the elastic constant remains the same. In our case, however, increasing $R$ will lead to the decreasing of $\varepsilon$ while the computation domain $\Omega$ remains the same. Their approach is more physical, while ours is more mathematical.
\end{remark}

\begin{remark}
 Before rescaling, the eigenvalues of $\tens{Q}$, $\lambda_i, i = 1,2,3$, take values in $[-\frac{1}{3}, \frac{2}{3}]$~\cite{majumdar2010equilibrium}. $\lambda_i = \frac{2}{3}$ corresponds to the case in which all LC molecules are pointing exactly at the same direction, whereas $\lambda_i = -\frac{1}{3}$ corresponds to the case in which LC molecules are completely compressed along the corresponding eigen-direction. After scaling, the eigenvalues of the scaled $\tens{Q}$ take value in $(\lambda_{min}, \lambda_{max})$, with $\lambda_{min} = -\frac{1}{3}\sqrt{\frac{27C^2}{2B^2}}$ and $\lambda_{max} = \frac{2}{3}\sqrt{\frac{27C^2}{2B^2}}$.
\end{remark}

The effective temperature $t$ appears only in the bulk energy term in the LdG. For $-\infty < t < 1$, nematic phase is energetically favored. Minimizing the bulk energy yields
\begin{equation}
 \tens{Q}^+ = s^+ \left( \bfmath{n}\bfmath{n} - \frac{\tens{I}}{3}\right),
 \label{eq_Qplus}
\end{equation}
where
\begin{equation}
 s^+ = \sqrt{\frac{3}{2}} \cdot \frac{3 + \sqrt{9 - 8t}}{4}.
 \label{eq_splus}
\end{equation}
Under certain boundary conditions, forcing $\tens{Q}$ everywhere to be of the form of Eq.~\eqref{eq_Qplus} will have to introduce singularities in $\bfmath{n}$, or defects. In order to reduce the total free-energy near the defects, $\tens{Q}$ may take the more general form of Eq.~\eqref{eq_Q}. Defect pattern, i.~e., the global positioning of singularities and the local profile near them is a delicate balance between the bulk, elastic and boundary energy. The study of defect pattern in LCs is important because: (i) Defects are the most visually striking feature of LC material and are closely related to its physical properties. (ii) Regions at or near defects challenge the limitation set by the models and are the ideal subject to study if we want to understand the relationship between different models.

A model system to study defect pattern is a spherical droplet of LCs with homeotropic anchoring condition at the boundary. All the three continuous models mentioned above admit the so-called \emph{radial hedgehog} solution, in which there exists one point defect with \emph{topological charge} 1 at the center of the ball~(Fig.~\ref{fig_bc1_three}(a)). For the LdG model, it has been shown both numerically and theoretically that the radial hedgehog solution is not stable for low temperature $t$, and the point defect will broaden into a \emph{disclination ring} (Fig.~\ref{fig_bc1_three}(b))~\cite{schopohl1988hedgehog, mkaddem2000fine, gartland1999instability, ignat2014stability}. The disclination ring is a symmetry breaking solution. Each point at the ring is a defect with topological charge +1/2, and the ring of defect is coated with a torus of biaxial region. As we mentioned earlier, the Oseen-Frank model can only admit isolated point defects, hence the disclination ring solution does not exist in the Oseen-Frank model. For the Ericksen's model, although it has been argued that it can predict the disclination ring solution~\cite{lin2001static}, the shape and stability of the ring may be quite different than that predicted by the LdG model because the Ericksen's model does not allow biaxiality. 

One can see that, even for the above simple model system, drastic difference in defect pattern exists among models. In order to gain a deeper understanding of defect patterns, including different types of defects and their transition, the global position and local profile of defects, and their parameter dependency, we study a spherical droplet of LCs subject to planar anchoring condition at the boundary. We numerically solve the LdG model with a spectral method based on Zernike polynomial expansion~\cite{zernike}. The high accuracy of this method allows us to capture the detailed configuration of defects. Based on our numerical results, we classify defects in the LdG model into five categories (see the end of Sec.~\ref{subsec_I}). Four of them involve disclination lines, suggesting that disclination lines are more energetically favored than point defects in the LdG model. In addition, we notice that disclination lines are always accompanied by biaxiality.

Given the importance of disclination lines in the LdG model, we systematically study a disk of nematic LCs as a model system of disclination lines. Assuming invariant of $\tens{Q}$ along the $z$-axis, a point defect in a 2-disk corresponds to a vertical disclination line of a cylinder. On the numerical side, we obtain three types of configurations for a variety of boundary conditions. The first type is stable only for high temperature and large $\varepsilon$. It has one single disclination line perpendicular to the center of the disk, the topological charge of which is determined by the boundary condition, with possible values $\pm k/2, k = 1, 2, 3, \cdots$. As the temperature and $\varepsilon$ decreases, a disclination line with $|k| > 1$ will quantize to $k$ separate $\pm 1/2$-disclination lines. This phenomena is consistent with a statement proved in~\cite{bauman2012analysis}. For certain boundary conditions, the system may admit a third type of solution, which is non-singular over the entire $\Omega$ and is also known as ``escaping in the third dimension"~\cite{sonnet1995alignment}. On the theoretical side, based on insights gained from numerical results, we obtain analytical expression of the profile of disclination lines. These profiles show how the defect in the center of the disk connect with the boundary through a biaxial region. Our results are similar to a class of special solutions for the LdG model reported in~\cite{fratta2014profiles}. Finally, to summarize the defect patterns in 3-ball and 2-disk we propose four conjectures. Together, these conjectures provide an integrated description of disclination lines --- from their global position to local profile. They also serve as important open questions for future research.

The rest of the paper is organized as follows. In Sec.~\ref{sec_II} we present our main numerical results. In Sec.~\ref{sec_III} the profiles of disclination lines are given analytically. In Sec.~\ref{sec_IV} a comparison between the LdG model and the Oseen-Frank model is made to highlight the fundamental difference between tensor and vector description of LCs. Finally, four conjectures of defect pattern are stated in Sec.~\ref{sec_V} alone with some open problems.

\section{Methods and results}
\label{sec_II}
First we give a brief description of the algorithm used in this paper. The goal is to find $\tens{Q}(\bfmath{x})$ that minimize the LdG free-energy in Eq.~\eqref{eq_totalenergy}, plus a penalty term that is to enforce the boundary condition (see below). We first expand $\tens{Q} \in \mycal{S}_0$ using orthogonal polynomials. Then we use BFGS algorithm~\cite{avriel2012nonlinear} to minimize the total energy iteratively and determine the expansion coefficients. This spectral method is particularly suitable for regular geometry shape such as the ball or disk considered here. Compared with finite difference or finite element algorithm, it can achieve high accuracy with a moderate number of variables. More detailed explanation of the algorithm is in the Appendix.

To visualize biaxiality, we follow~\cite{mkaddem2000fine} and define
\begin{equation}
\beta = 1 - 6\frac{(\mathrm{tr}\tens{Q}^3)^2}{(\mathrm{tr}\tens{Q}^2)^3}.
\label{eq_beta}
\end{equation}
For uniaxial $\tens{Q}$ $\beta = 0$ while for biaxial $\tens{Q}$ $\beta \ne 0$. 

To visualize the tensor field, we define
\begin{equation*}
 \tens{D} = \frac{\tens{Q}^{\mathrm{diag}} - \lambda_{min}\tens{I}}{\lambda_{max} - \lambda_{min}} = \left( \begin{array}{ccc}
d_1 & 0 & 0 \\
0 & d_2 & 0 \\
0 & 0 & d_3 \end{array} \right),
\end{equation*}
where $\tens{Q}^{\mathrm{diag}}$ is the diagonalized matrix of $\tens{Q}$ and the eigenvalues of $\tens{D}$ satisfy $d_1 \ge d_2 \ge d_3 \ge 0$ and $d_1 + d_2 + d_3 = 1$. We use an ellipsoid whose three semi-principle axes lie in the eigenvectors of $\tens{Q}$ with length equal to the corresponding eigenvalues. In this representation, an isotropic $\tens{Q}$ is a ball and a uniaxial $\tens{Q}$ with positive (negative) $s$ is a prolate (oblate). 

To visualize defects, following~\cite{callan2006simulation} we define 
\begin{equation}
c_l = d_1 - d_2, \ \ c_p = 2(d_2 - d_3), \ \ c_s = 3d_3.
\label{eq_cl}
\end{equation}
$c_l, c_p$ and $c_s$ satisfy the properties
$$
0 \le c_l, c_p, c_s \le 1,
$$
and
$$
 c_l + c_p + c_s = 1.
$$
At defect, $c_l=0$, so the iso-surface of $c_l = \delta$ for a small positive constant $\delta$ is an indication of where the defect is. 

\begin{remark}
In vector models of LC such as the Oseen-Frank model and the Ericksen's model, defects are defined as discontinuity in $\bfmath{n}$. For tensor model like the LdG, it is not straightforward to define defect because the map from a tensor $\tens{Q} \in \mycal{S}_0$ to a vector $\bfmath{n}' \in \mathbb{S}^2$ can be ambiguous. For example, one can choose $\bfmath{n}'$ as the eigenvector corresponding to the largest eigenvalue of $\tens{Q}$~\cite{nguyen2010refined}. But when $\tens{Q}$ is oblate, this $\bfmath{n}'$ contradict with the $\bfmath{n}$ defined in Eq.~\eqref{eq_uset}. Efforts have been made in rigorously defining defect for a tensor field~\cite{Paolo1997}. However, it is not the focus of this work and all the defects we meet are relatively easy to be identified.
\label{rmk1}
\end{remark}

\subsection{Ball}
\label{subsec_I}
\begin{figure}
\centering
\includegraphics[width=\textwidth]{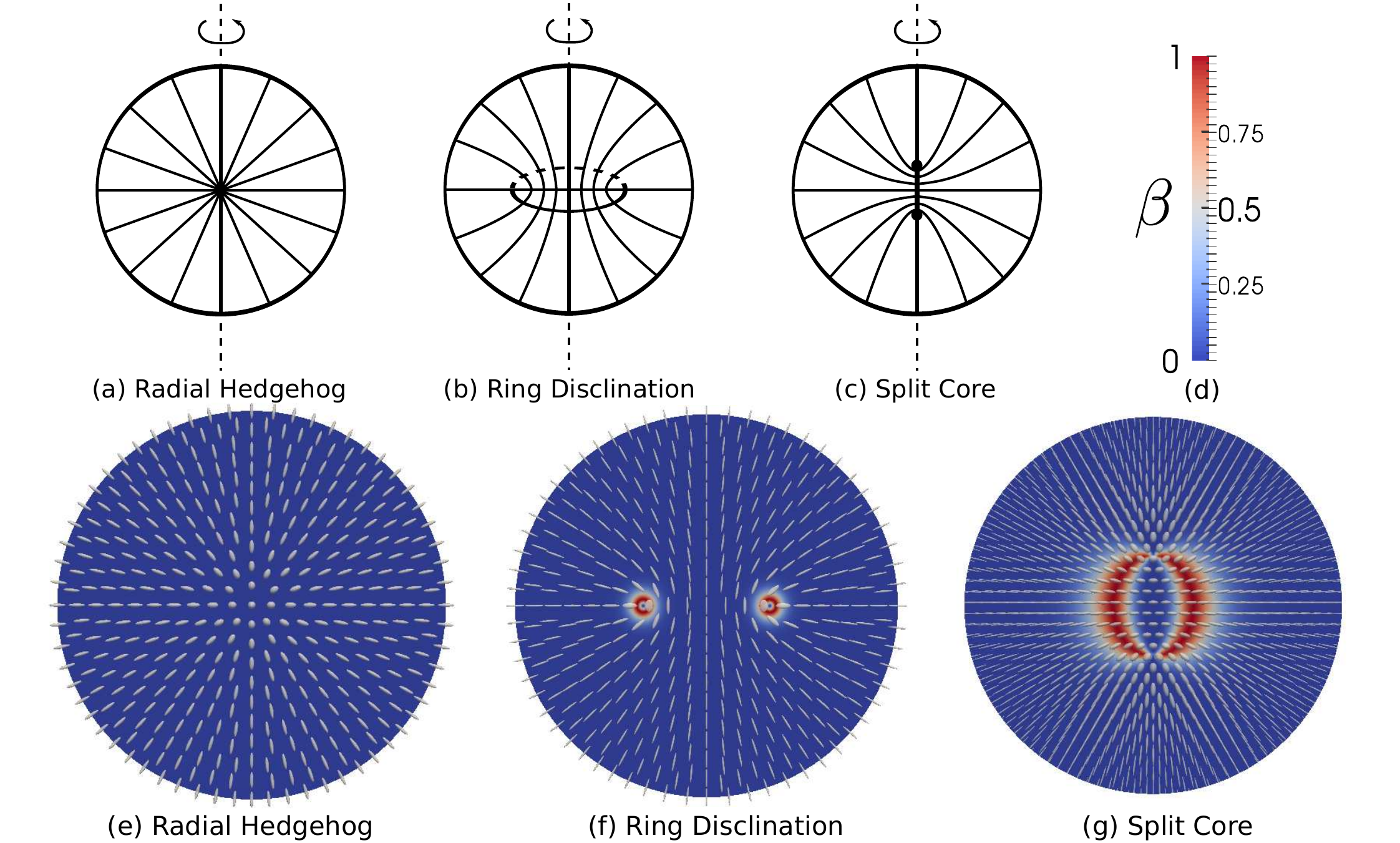}
\caption{Three possible configurations under the strong radial anchoring condition. (a-c) Qualitative rendering of the alignment direction of the radial hedgehog, ring disclination and split core on the $xz$-plane (d) Color bar for $\beta$ shown in (e-g), with red indicates biaxial and blue indicates uniaxial. (e-g) $\beta$ (represented by color) and $\tens{Q}$-tensor (represented by ellipsoid glyph) from numerical simulation. In all the three cases $\varepsilon = 0.2$, and the temperatures are (e) $t=-2$, (f) $t=-6$ and (g) $t=-12$. (e) and (f) show the whole computational domain, while (g) only shows a zoom-in view of radius $=0.3$ for a better resolution of the defect pattern in the center of the ball.}
\label{fig_bc1_three}
\end{figure}

First we consider the \emph{strong radial anchoring condition}. 
The surface free-energy density is given by
$
f_s(\bfmath{x}) = \omega (\tens{Q}_{ij}(\bfmath{x}) - \tens{Q}^+_{ij}(\bfmath{x}))^2,
$
for $\bfmath{x} \in \partial \Omega$. Here $\tens{Q}^{+}(\bfmath{x})$ satisfies Eq.~\eqref{eq_Qplus} (with $\bfmath{n}$ replaced by $\bfmath{x}$). $\omega$ is a constant that controls the relative strength of anchoring. We obtain three different configurations as shown in Fig.~\ref{fig_bc1_three}. These are the \emph{radial hedgehog}, \emph{ring disclination} and \emph{split core} solutions obtained in~\cite{mkaddem2000fine} by assuming rotational symmetry around the $z$-axis. Here we recover these solutions in a full three-dimensional computation. It was guessed that the split core solution is not stable after removing the rotational symmetry assumption~\cite{mkaddem2000fine}. Here we find that, for parameters within a certain region, the split core solution is stable when subject to a moderate level of perturbation.

\begin{figure}
\centering
\includegraphics[width=.8\textwidth]{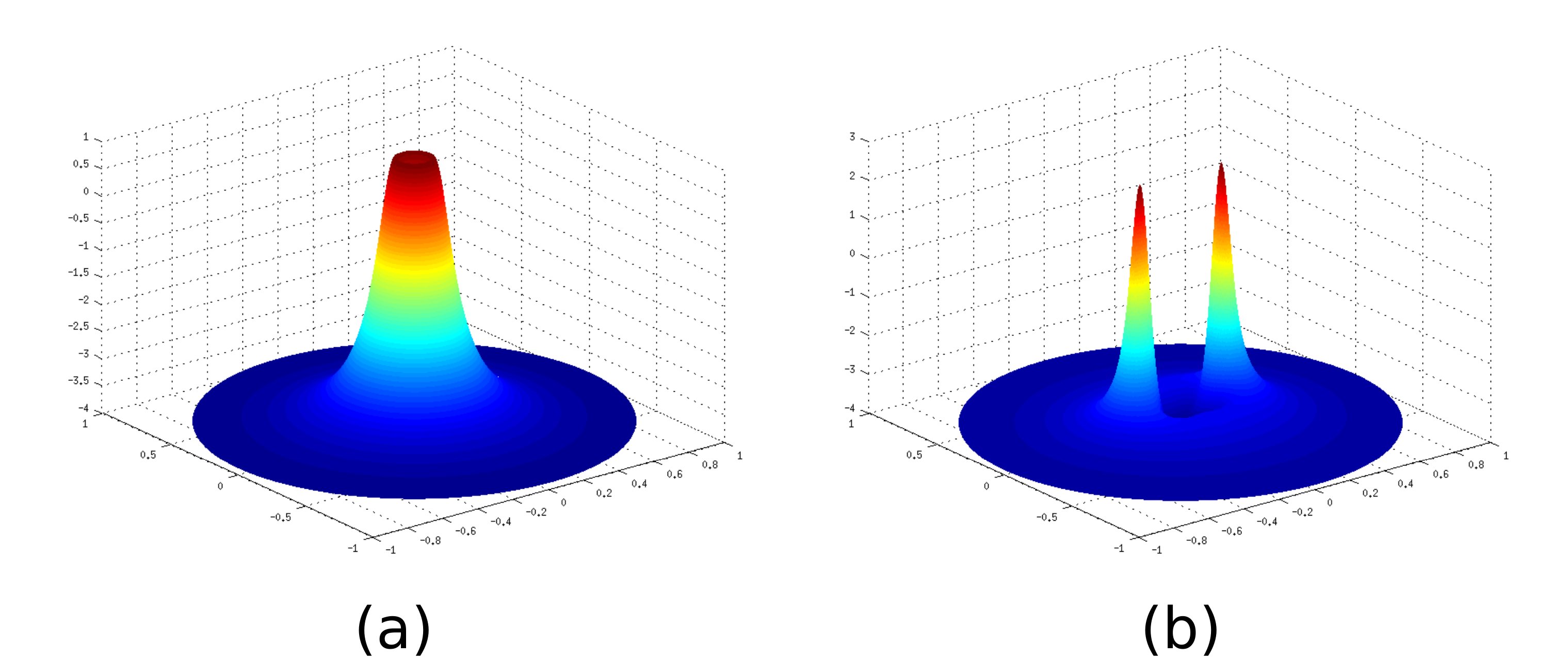}
\caption{Energy concentration near defects. The plot region corresponds to the $xz$-plane shown in Fig.~\ref{fig_bc1_three}. Both the height and color correspond to the total free-energy density. (a) radial hedgehog. $t=-2$, $\varepsilon=0.2$. (b) disclination ring. $t=-6$, $\varepsilon=0.2$.}
\label{fig_bc1_energy}
\end{figure}

In the radial hedgehog solution $\tens{Q}$ is uniaxial everywhere (Fig.~\ref{fig_bc1_three} (a) and (e)). The center of the ball is the only point defect (with topological charge +1). For small $t$ and $\varepsilon$, this point defect broadens into a disclination ring (Fig.~\ref{fig_bc1_three} (b) and (f)). The ring is composed of point defects with charge +1/2. Detailed study of the relation of the ring structure on $t$ and $\varepsilon$ is documented in~\cite{mkaddem2000fine}. The disclination ring is a symmetrical-breaking configuration. Around the ring a torus of strong biaxial region ($\beta \sim 1$) exists. The split core solution contains a short +1 disclination line in the center (Fig.~\ref{fig_bc1_three} (c) and (g)), with two isotropic points at both ends. It is also shelled by a strong biaxial region. As we mentioned earlier, both defects and biaxiality are structures that are not energetically favored. As a result, their existence will raise the local energy density. Fig.~\ref{fig_bc1_energy} shows that the total free-energy are concentrated near the central point defect in the radial hedgehog solution. Note that there is a small dip in the energy landscape at the center and the maximum of energy density is reached at a small distance away from the point defect. In the disclination ring solution, the total energy is concentrated near the biaxial torus, with maximum reached right at the disclination ring. Between these two ways of distributing energy, the second one is more economic (in the sense that it lowers the total energy) at low temperature. We will come back to this point in the Discussion. 

\begin{remark}
It was proposed in~\cite{lin2001static} that disclination ring configuration can be predicted by the Ericksen's model. We try to verify it numerically by imposing uniaxial constraint over $\tens{Q}$ (forcing $\beta=0$ by introducing a penalty term). The rational behind this procedure is that the LdG model with $\tens{Q}$ constrained in the uniaxial region is equivalent to the Ericksen's model. Surprisingly, within the parameter range we tested, the only stable uniaxial solution we get is the radial hedgehog. The inconsistence between our numerical results and theoretical reasoning made in~\cite{lin2001static} might be caused by the limited parameter region our method can handle.
\end{remark}

\begin{figure}
\centering
\includegraphics[width=.8\textwidth]{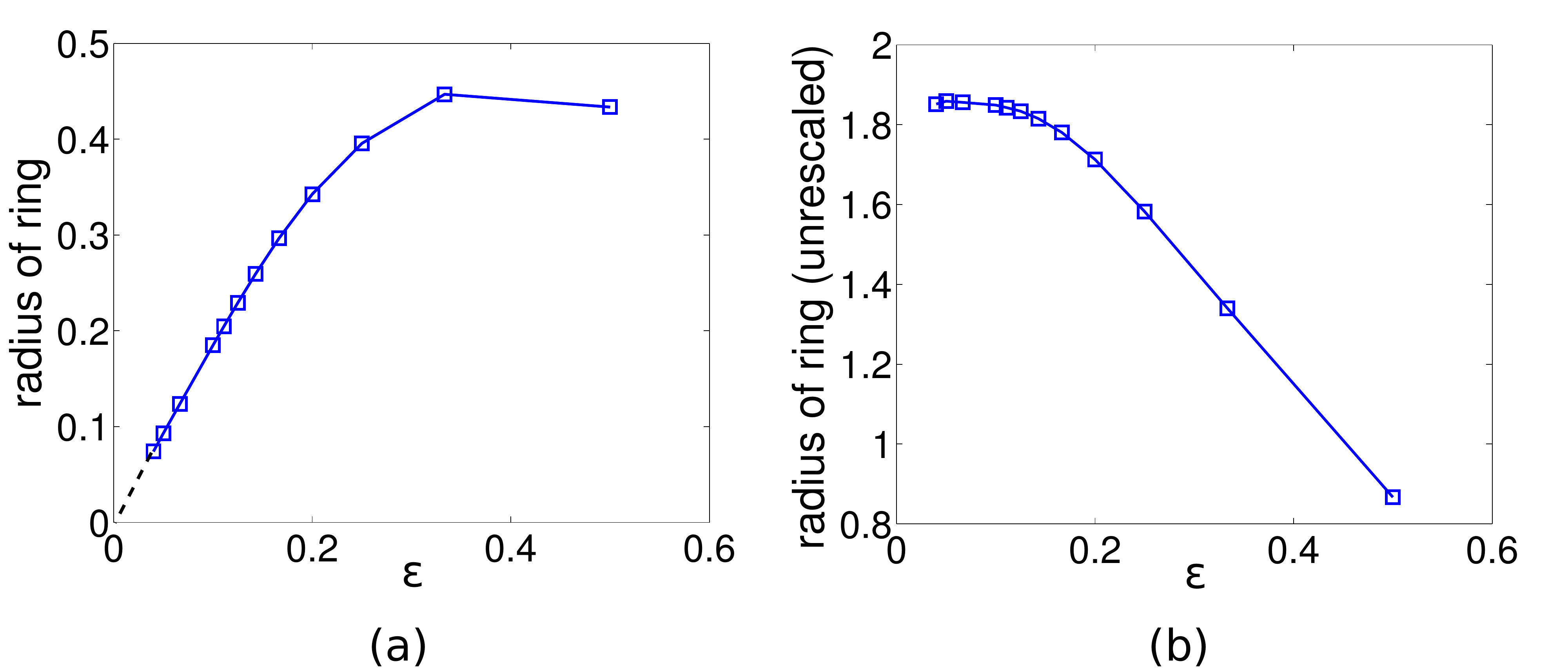}
\caption{(a) Radius of the ring as a function of $\varepsilon$ under the strong radial anchoring condition. $t = -5$. The black-dashed line is produce based on the perception that the radius approaches to 0 in the limit of $\varepsilon \rightarrow 0$. (b) Same results with (a) but the radius of the radius is measured by the characteristic length $r_{ring}' = r_{ring} R / \xi_0$.}
\label{fig_bc1_rr}
\end{figure}

It was proved in~\cite{majumdar2010landau} that, for strong radial boundary and sufficiently low temperature, the global minimizer of the LdG energy will converge strongly to that predicted by the Oseen-Frank theory, in the limit of $\varepsilon \rightarrow 0$. In particular, it means the disclination ring will converge to the radial hedgehog as $\varepsilon \rightarrow 0$. To verify this result, we measure the radius of the ring, $r_{ring}$, defined as the distance between the center of the ball to a point on the ring, for fixed $t$ and different $\varepsilon$. Fig.~\ref{fig_bc1_rr} (a) shows as $\varepsilon$ gets smaller, the radius of the ring decreases, consistent with the above statement. On the other hand, if measured in the characteristic length $r'_{ring} := r_{ring} R/ \xi_0 = r_{ring}/\varepsilon$, the \emph{actual} radius of the ring $r'_{ring}$ seems to approach to a constant as $\varepsilon \rightarrow 0$ (Fig.~\ref{fig_bc1_rr} (b)). The observation that $r'_{ring}$ has a finite limit was also made in~\cite{mkaddem2000fine} but, to our best knowledge, a mathematical proof is still missing. To summarize, the actual size of the disclination ring configuration approaches to a finite size as the radius of the ball $R \rightarrow \infty$, which is determined by the material properties and temperature only. It is the rescaling which maps a ball with infinitely large radius to a unit ball that leads the \emph{rescaled} radius of the ring to zero. 

\begin{figure}
\centering
\includegraphics[width=.8\textwidth]{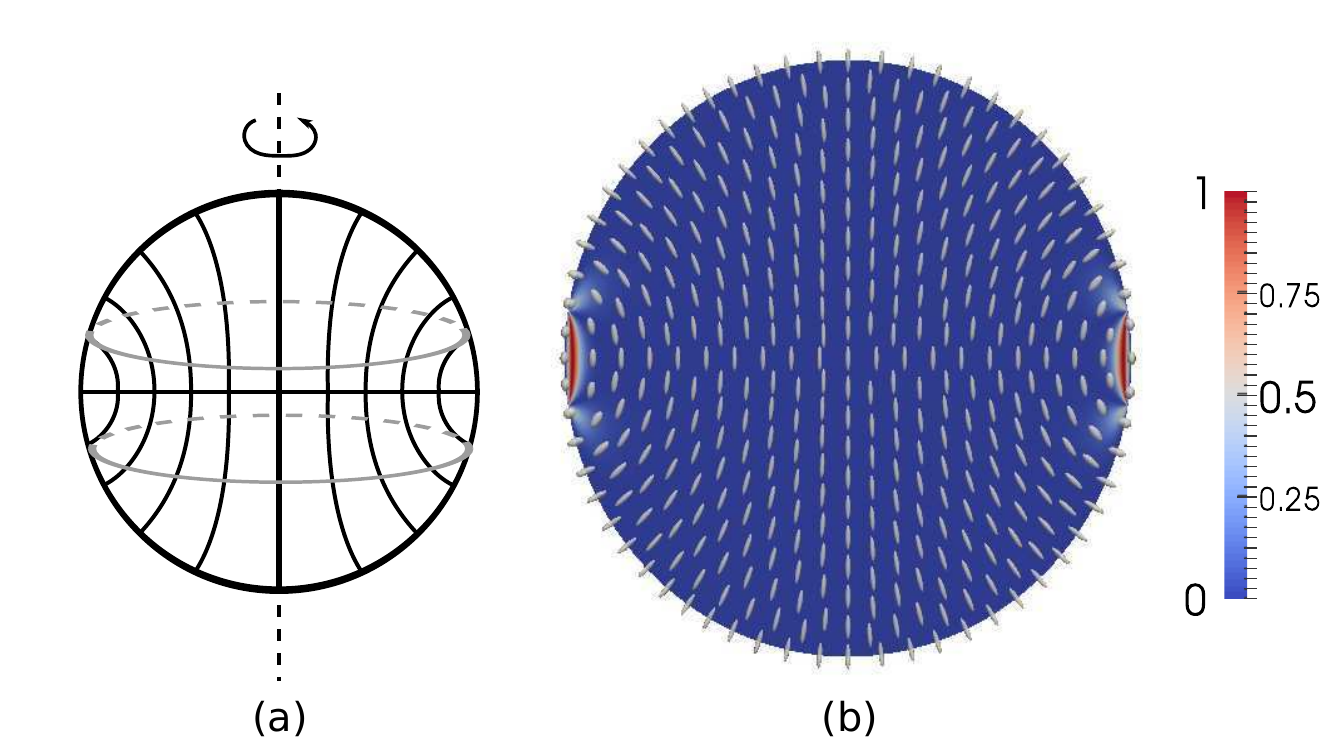}
\caption{Solution under the relaxed radial anchoring condition. (a) Qualitative rendering of the alignment direct field. Isotropic points form two parallel closed loops on the sphere (gray curve). (b) $\beta$ (represented by color) and $\tens{Q}$-tensor (represented by ellipsoid glyph) from numerical simulation. Parameters used are $\varepsilon = 0.2$ and $t=-2$.}
\label{fig_bc2}
\end{figure}

The strong radial anchoring condition requires $s=s^+$ to be fixed at the boundary. Next, we relax this requirement and allow $s(\bfmath{x})$ to be a free scalar function on $\partial{\Omega}$, i.~e.,
\begin{equation*}
 \tens{Q}(\bfmath{x}) = s(\bfmath{x}) \left( \bfmath{x}\bfmath{x} - \frac{\tens{I}}{3}\right), 
 \ \ \bfmath{x} \in \partial \Omega.
\end{equation*}
We call it the \emph{relaxed radial anchoring condition}. Besides the radial hedgehog, disclination ring and split core configurations, we obtain an additional stable solution for this boundary condition as shown in Fig.~\ref{fig_bc2}. This solution was also reported in~\cite{porenta2011effect}. In it, two rings of isotropic points form on the sphere. Between the two rings, on the surface $Q$ is uniaxial (as required by the boundary condition ) and oblate ($s<0$). Inside there is a strong biaxial region close to the surface.

Next we consider the more complex planar boundary condition. Due to the topological constraint imposed by the spherical surface, it is no longer possible to restrict $\tens{Q}$ in the form of Eq.~\eqref{eq_Qplus} everywhere on the boundary without introducing any defect (a result known as the ``hairlyball theorem''~\cite{hairlyball}). Instead, we demand $\tens{Q}(\bfmath{x}) \in \mycal{C}$ for $\bfmath{x} \in \partial \Omega$, with
\begin{equation*}
 \mycal{C} = \{ \tens{Q} \in \mycal{S}_0: \tens{Q} \bfmath{\nu} = \lambda_\nu \bfmath{\nu}\},
\end{equation*}
where $\bfmath{\nu}$ is the normal direction of the surface and $\lambda_{min} \le \lambda_\nu < 0$ is a constant. $\bfmath{\nu}$ measures the strength of compression imposed on the LC molecules at the boundary along the normal direction ($\lambda_\nu > 0$ corresponds to extension rather than compression). In particular we choose $\lambda_\nu = s^+\lambda_{min}$ with $s^+$ given by Eq.~\eqref{eq_splus} but other choices of $\lambda_\nu$ can be made here as well. The boundary energy density is given by
\begin{equation}
f_s = \omega \left\| \left(\tens{Q} + \lambda_\nu \tens{I}\right)\bfmath{x} \right\|^2,  \ \ \bfmath{x} \in \partial \Omega.
\label{eq_bc3}
\end{equation}
Here $\| \cdot \|$ is the second-order vector norm.
This boundary condition is a special case of the one used in~\cite{tasinkevych2012liquid} (with $W_2 = 0$ in Eq. (6) of that paper).

\begin{figure}
\centering
\includegraphics[width=\textwidth]{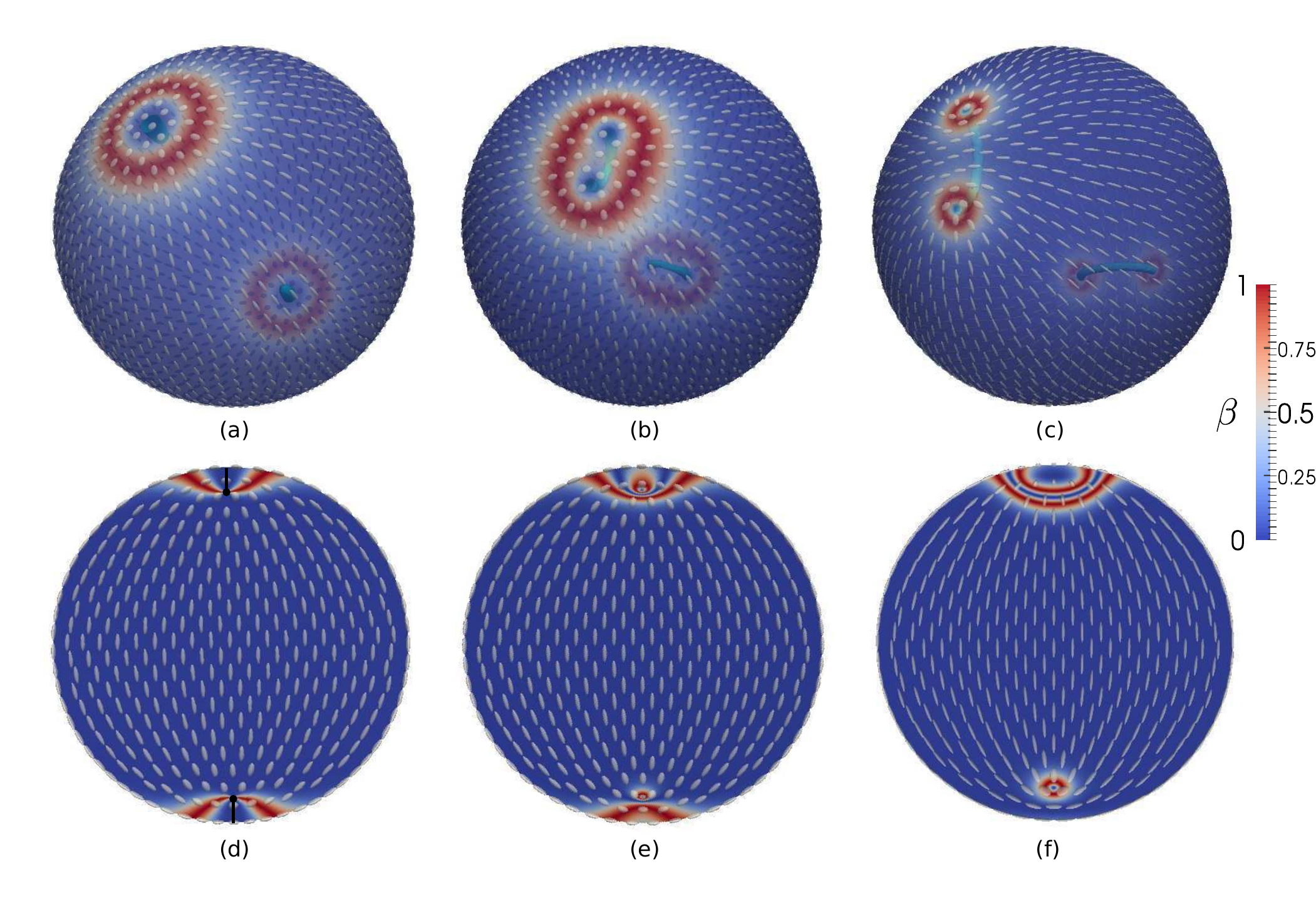}
\caption{Three stable solutions for the planar anchoring condition for fixed $\varepsilon=0.25$ and different $t$ (a and d: $t=-1$; b and e: $t=-1.1$; c and f: $t=-4$). (a-c) three-dimensional view. The ellipsoid represent the $\tens{Q}$-tensor on the surface. Color corresponds to $\beta$, ranging from 0 (blue) to 1 (red). The tubes inside the ball in (a)-(c) are the iso-surface of $c_l$, with values equal to (a): $c_l = 0.01$; (b): $c_l = 0.03$; (c): $c_l = 0.1$). (d-f) Sliced view to show the inside of the ball. The cutting plane is determined by the ball center and a pair of defect points on the surface. $\beta$ is shown in color and $\tens{Q}$-tensor is represented by ellipsoid glyph. The thick black lines in (d) represent two segments of +1 disclination lines.}
\label{fig_bc3_three}
\end{figure}

\begin{figure}
\centering
\includegraphics[width=.8\textwidth]{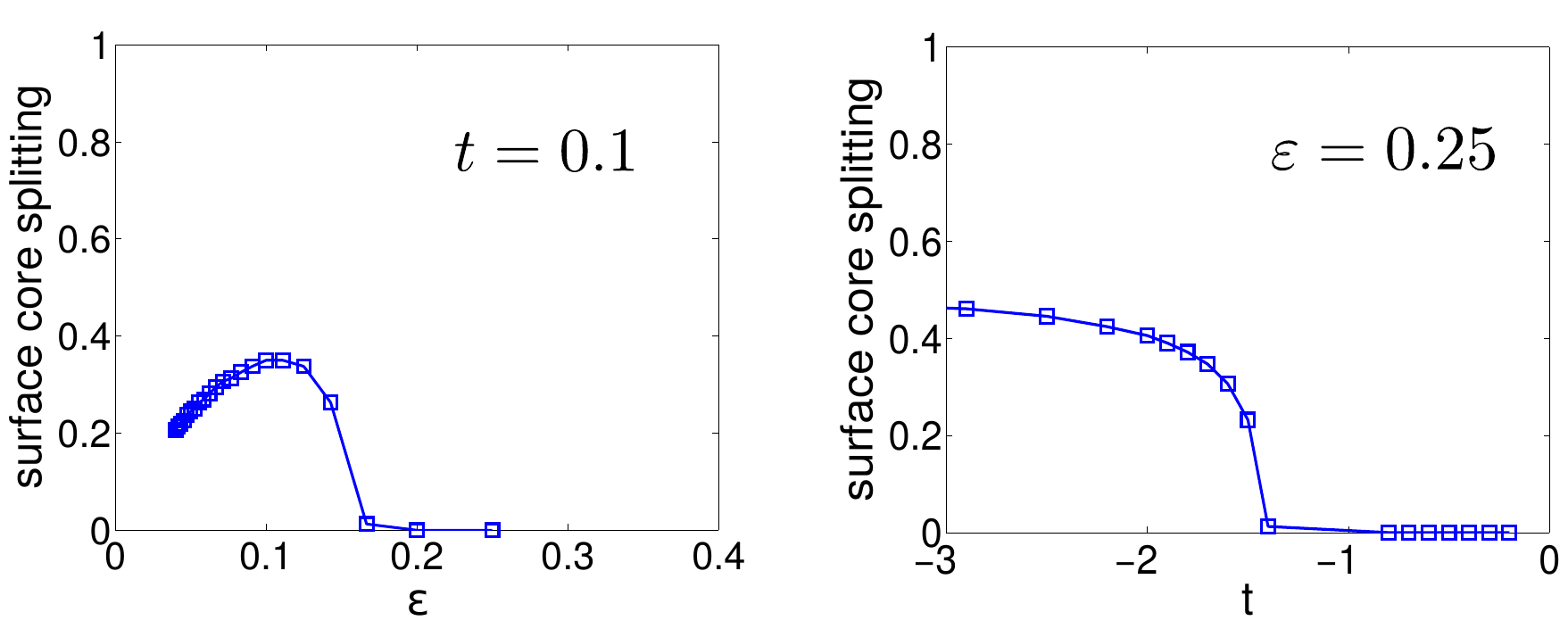}
\caption{Distance between two coupling +1/2 point defects on surface (see Fig.~\ref{fig_bc3_three}) as a function of $\varepsilon$ and $t$.}
\label{fig_rt}
\end{figure}

Fig.~\ref{fig_bc3_three} illustrates the defect pattern under the planar anchoring condition Eq.~\eqref{eq_bc3} for fixed $\varepsilon=0.25$ and different $t$. First we only look at the defect pattern on the surface. For $t=-1$, two +1 point defects form at two poles (Fig.~\ref{fig_bc3_three} (a)). Around each defect point there is ring of biaxial region. As temperature decreases, the point defect on the surface will split into two point defects with topological charge +1/2. During this transition, the biaxial ring will shrink in one direction and elongate in the other, a process similar to a cell dividing into two daughter cells on a culture plate. Fig.~\ref{fig_bc3_three} (b) shows an intermediate step ($t=-1.1$) in which the biaxial region has not separated, whereas in (c) the two newly developed biaxial rings are fully separated ($t=-4$). In Fig.~\ref{fig_bc3_three} (c), the four +1/2 point defects on the sphere form the vertices of a tetrahedron. This conformation is similar with the \emph{tennis ball} solution~\cite{nelson2002toward, zhang2012onsager, cheng2013tensor} obtained for LC-sphere (one sheet of LC molecules confined on a spherical surface, no LCs inside the ball). For LC-sphere, the four point defects form a regular tetrahedron. Here, the tetrahedron is not a regular one due to the influence of the LC bulk inside the ball. We measure the distance $d$ between two neighboring 1/2-point defects for different $t$ and $\varepsilon$. Similar with the radius of the disclination ring for the homeotropic anchoring condition, it appears that $d$ approaches to zero as $\varepsilon \rightarrow 0$ (Fig.~\ref{fig_rt}), and to a finite constant if measured in characteristic length (results not shown here). Finally we note that, a similar transitional process in which a +1 point defect splits into two +1/2 point defects on the surface has been studied in~\cite{tasinkevych2012liquid} for a solid spherical body immersed in nematic LC host. The three states corresponding to Fig.~\ref{fig_bc3_three} (a), (b) and (c) was named as \emph{single core}, \emph{double core} and \emph{split core} in~\cite{tasinkevych2012liquid} and we will follow these names below. While our results are qualitative similar with theirs, the confinement of LC inside the ball poses different constraint to the system.

Now we examine defect pattern inside the ball. In Fig.~\ref{fig_bc3_three} (a-c) the isosurface of $c_l$ defined in Eq.~\eqref{eq_cl} is plotted to encapsulate the disclination lines. (d-f) show $\beta$ and $\tens{Q}$ inside the ball. We can see that, the above mentioned point defects on the surface are in fact the intersection between disclination lines developed inside the ball with the spherical surface. The single core solution has two segments of  disclination with topological charge +1 (indicated by the thick lines in (d)). One end of the disclination line is isotropic and buried inside the LC ball while the other end connects the surface. As temperature decreases, the +1 disclination will split into a +1/2-disclination with both ends open at the surface.

\begin{figure}
\centering
\includegraphics[width=.4\textwidth]{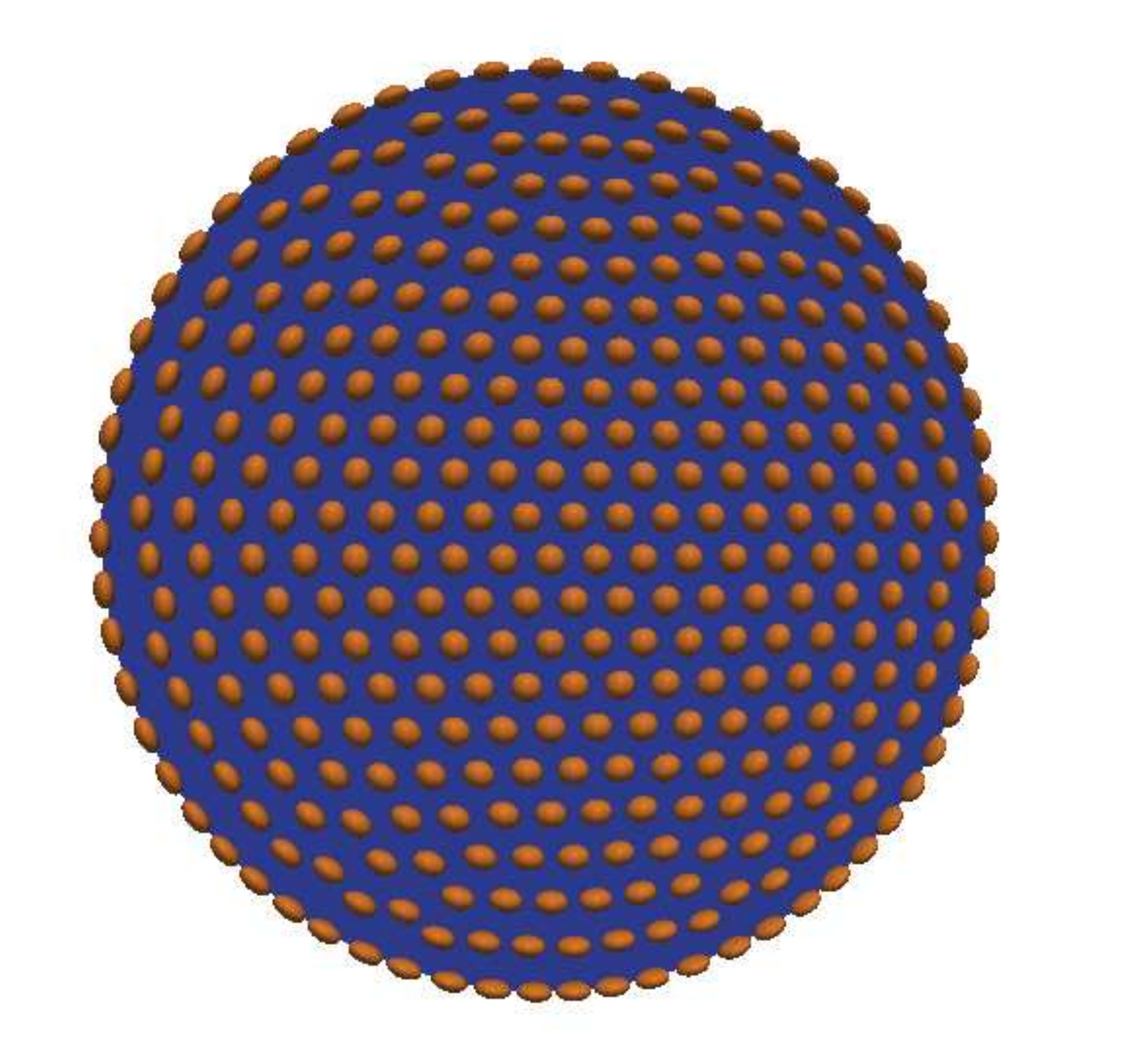}
\caption{A uniaxial solution for planar anchoring condition. $\beta=0$ and $\tens{Q}$ is oblate everywhere. $t=0.5$, $\varepsilon = 1$.}
\label{fig_bc3_flat}
\end{figure}

Besides the three solutions in Fig.~\ref{fig_bc3_three}, we found two other meta-stable solutions. The first one has a structure similar to the tennis ball configuration, but with one hemisphere rotated by $\pi/2$ around the $z$-axis so that the four +1/2 point defects on the surface lie on one big circle. We call this solution \emph{rectangle}. Its free-energy is higher than the tennis ball configuration. Another meta-stable state is shown in Fig.~\ref{fig_bc3_flat}. It only exists for large $t$ and $\varepsilon$. Like the radial hedgehog solution, $\tens{Q}$ in this configuration is uniaxial everywhere and satisfies radial symmetry, except it is oblate rather than prolate. For this configuration, if the tensor-field $\tens{Q}$ is mapped to a vector field $\bfmath{n}$ according to Eq.~\eqref{eq_uset}, one will obtain a singularity in the center of the ball, just like the radial hedgehog solution.

As a summary of the numerical results on the three-dimensional ball, we observe point defects and disclination lines for different anchoring conditions. It seems that disclination lines is more commonly found within the LdG model compared with point defects. Phenomenologically, disclination lines can be classified into four types:
\begin{enumerate}
 \item Disclination line form closed loop inside or on the surface of the ball (Fig.~\ref{fig_bc1_three}(f) and Fig.~\ref{fig_bc2}).
 \item Both ends of the disclination line submerged in the LC bulk (Fig.~\ref{fig_bc1_three}(g)).
 \item One end of the disclination line submerged in the LC bulk while the other end connects to the surface (Fig.~\ref{fig_bc3_three}(d)).
 \item Both ends of the disclination line connect to the surface (Fig.~\ref{fig_bc3_three}(e) and (f)).
\end{enumerate}
One feature that is shared by all disclination lines in the LdG model is that they are always accompanied by regions with strong biaxiality. In fact, the bulk energy $f_b$ does not favor $\tens{Q}$ that is biaxial. The fact that biaxiality is closely related to defects suggests that defect pattern is a subtle balance between the elastic energy and topological constraint. 

\subsection{Disk}
\label{sec_2dresult}
Disk is an ideal system to study the profile of disclination lines. A point defect in a disk $\Omega(x, y)$ is a vertical disclination line along the $z$-direction. Consider a unit disk $\Omega = D_1$, for different boundary conditions, we find $\tens{Q}(x, y) = \tens{Q}(r\cos \phi, r \sin \phi)\in \mycal{S}_0$ for $0 \le r \le 1, 0 \le \phi < 2\pi$ that minimize the LdG energy Eq.~\eqref{eq_totalenergy}.

First we consider the boundary condition
\begin{equation}
 \tens{Q}(\cos \phi, \sin \phi) = s^+(\bfmath{n}\bfmath{n} - \frac{\tens{I}}{3}),
 \label{eq_bc2d}
\end{equation}
with $\bfmath{n} = \left(\cos\frac{k}{2}\phi,\sin\frac{k}{2}\phi, 0\right)$, $k=\pm 1, \pm 2, \cdots$. $s^+$ is given by Eq.~\eqref{eq_splus}. Under this condition, $\bfmath{n}$ at the boundary always lies in the $xy$-plane. Traveling along the circle of $r=1$ rotates $\bfmath{n}$ by an angle of $k\pi$ (positive angle means counter-clockwise and negative angle means clockwise). 

\begin{figure}
\centering
  \includegraphics[width=0.6\textwidth]{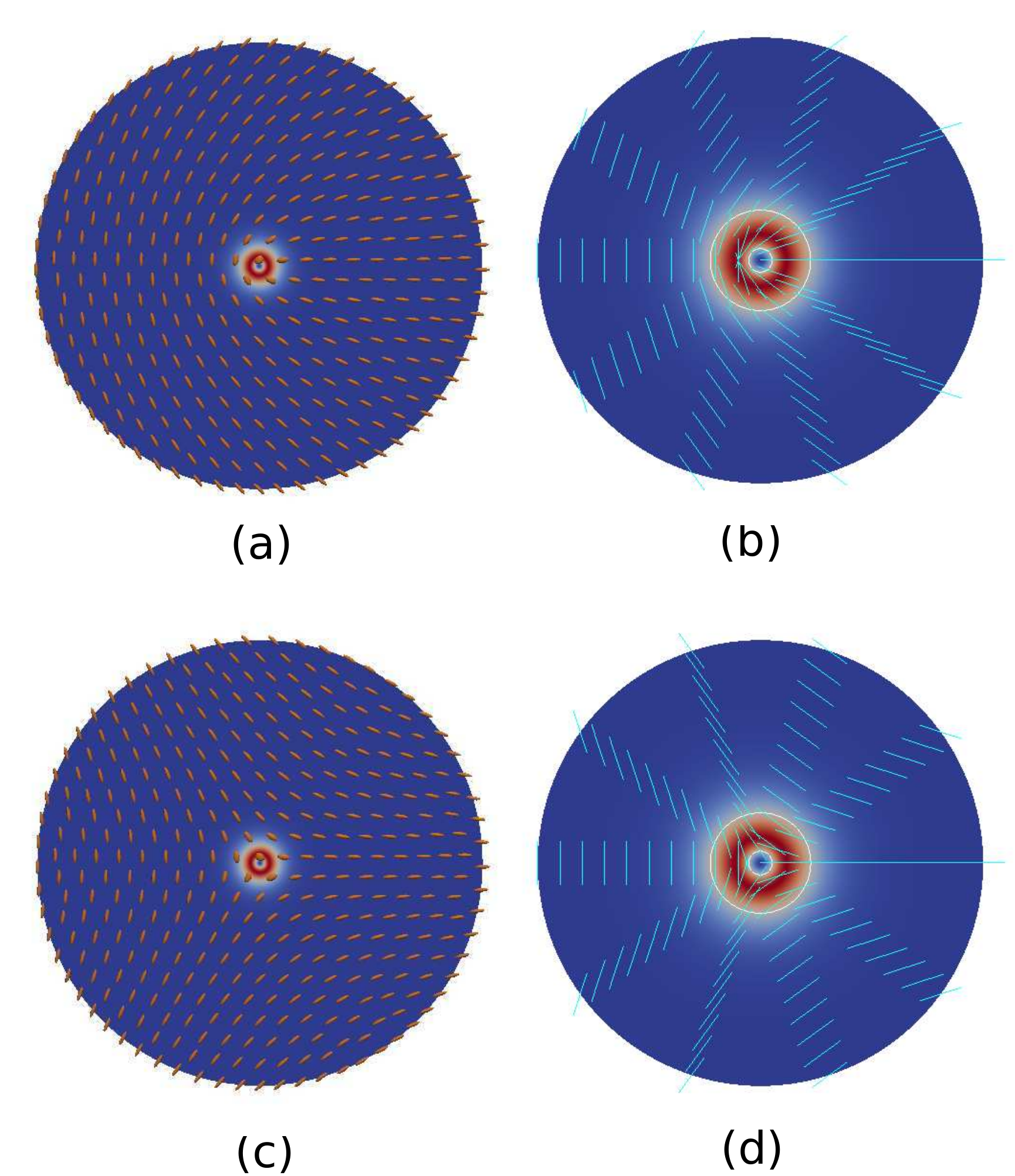}
 \caption{Solutions for $k=\pm 1$ under the boundary condition Eq.~\eqref{eq_bc2d}. $k=1$ in (a, b) and $k=-1$ in (c, d). $\beta$ is shown in color with red corresponds to biaxial and blue uniaxial. Ellipsoids represent the $\tens{Q}$-tensor. Parameters used are: (a, c) $t=-1$, $\varepsilon=0.2$. (b, d) $t=-0.1$, $\varepsilon=0.5$. In (b, d), the solid lines represent the eigenvectors corresponds to the largest eigenvalue of $\tens{Q}$, and the white circles are the contours for $\beta=0.5$.}
 \label{fig_2d1}
\end{figure}

For $k = 1$ we obtain solution shown in Fig.~\ref{fig_2d1} (a), (b). In the center of the disk, there is a +1/2 point defect, surrounded by a biaxial ring. For the same parameters, the solution for $k=-1$ has the same eigenvalues with that of $k=1$, only the eigenvectors are rotated, resulting a -1/2 point defect in the center (Fig.~\ref{fig_2d1} (c), (d)).

\begin{figure}
\centering
\resizebox{\textwidth}{!}{\includegraphics{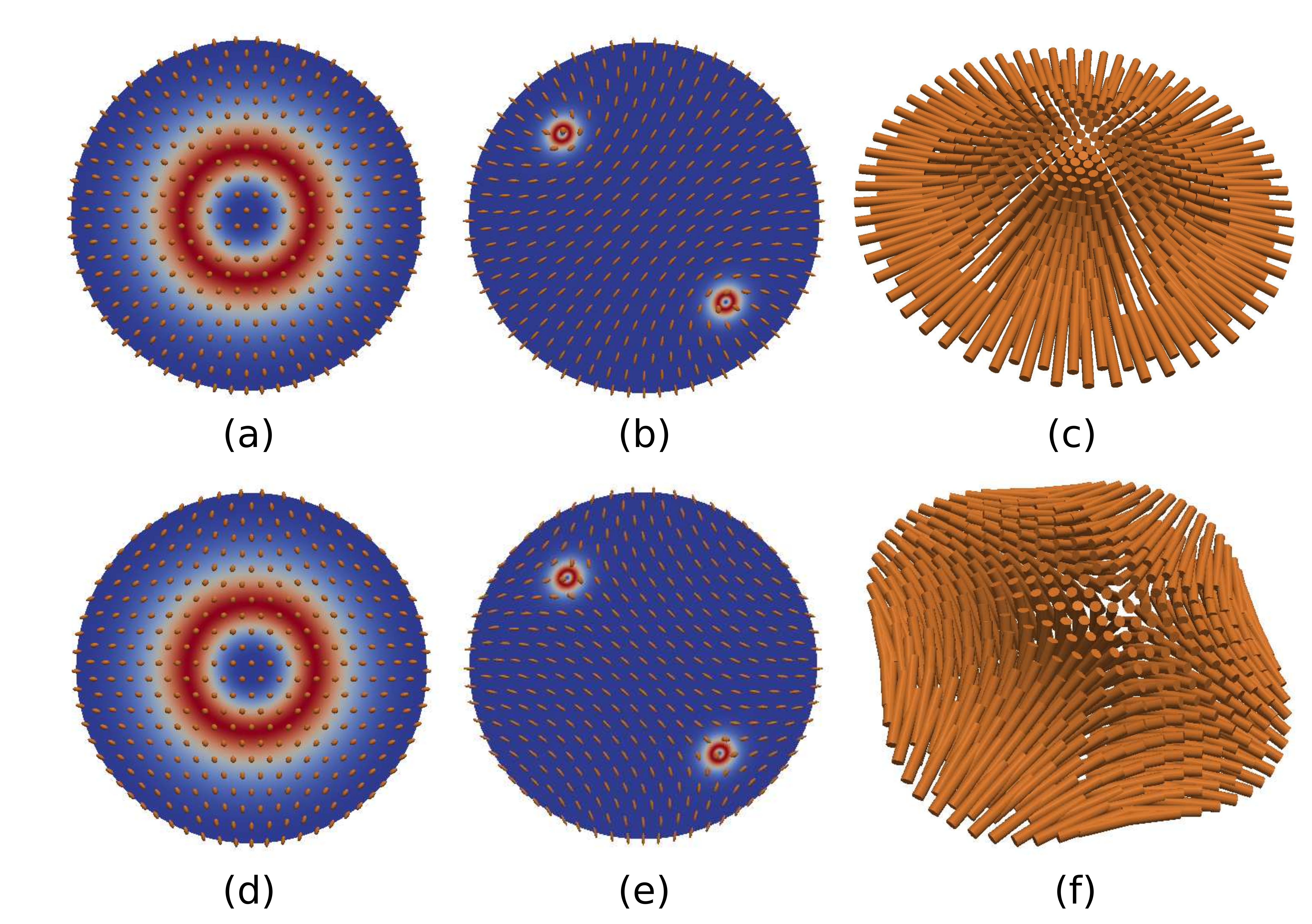}}
 \caption{Solutions for $k = 2$ (a-c) and $k=-2$ (d-f). $\beta$ is shown in color with red corresponds to biaxial and blue uniaxial. Ellipsoids represent the $\tens{Q}$-tensor. Golden solid bars in (c) and (f) represent the eigendirection corresponding to the largest eigenvalue. (a) planar radial, parameters used are $t=1$, $\varepsilon=0.5$. (b) planar polar, $t=-1$, $\varepsilon=0.2$. (c) escape radial, $t=-6$, $\varepsilon=0.2$. (c) $t=1$, $\varepsilon=0.5$. (d) $t=-1$, $\varepsilon=0.2$. (f) $t=-6$, $\varepsilon=0.2$.}
 \label{fig_2d2}
\end{figure}

\begin{figure}
\centering
  \includegraphics[width=0.5\textwidth]{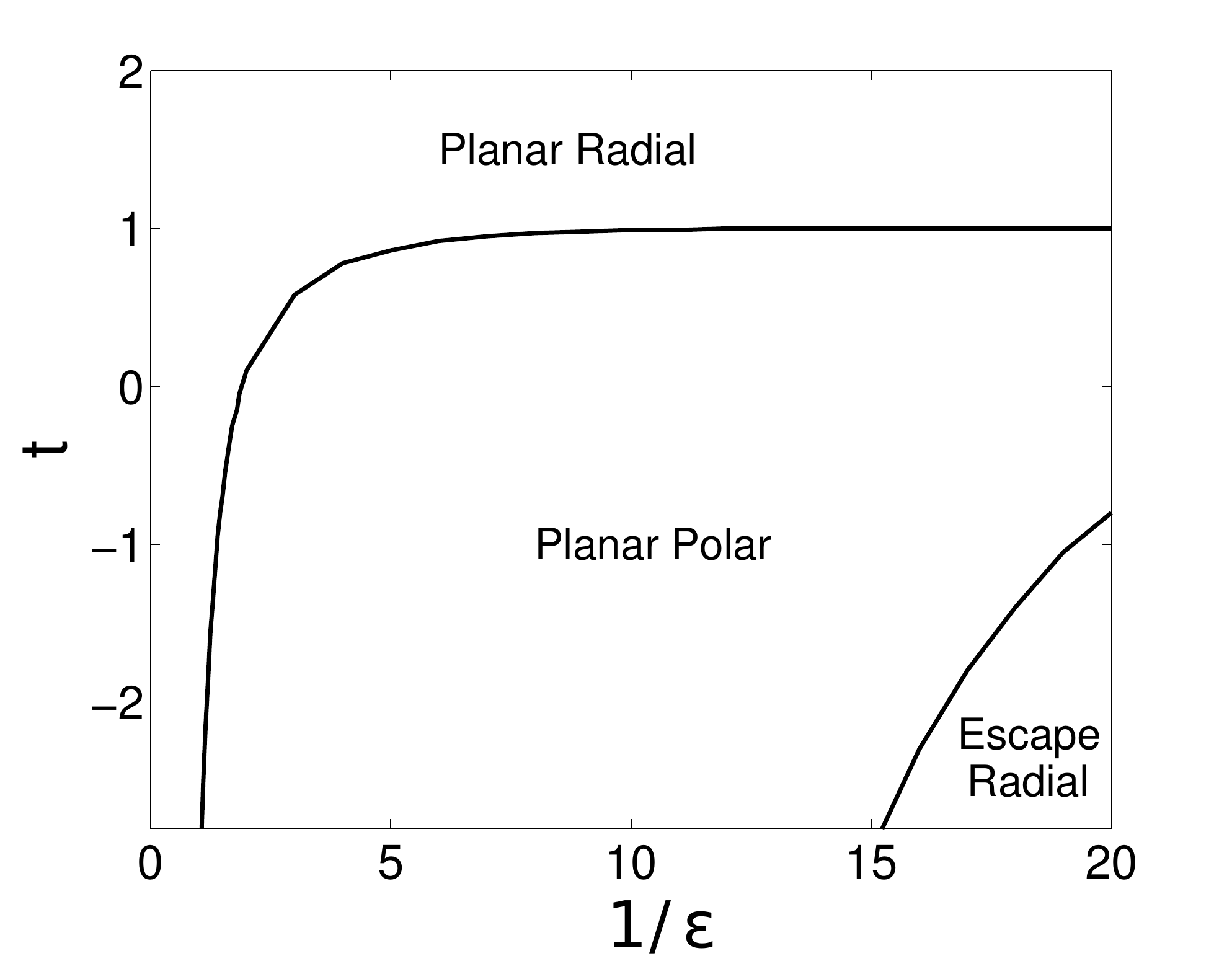}
 \caption{Phase diagram of the planar radial, planar polar and escape radial configurations for $k=2$. The partition is based on the lowest energy of the three.}
 \label{fig_2d_phasek2}
\end{figure}

The case for $k=2$ has been studied numerically in~\cite{allender1991determination, sonnet1995alignment, kralj1999}. Three possible configurations exist, which are the \emph{planar radial} (Fig.~\ref{fig_2d2}a), \emph{planar polar} (b), and \emph{escape radial} (c). The planar radial configuration exists for high temperature and large $\varepsilon$. For low temperature the planar polar configuration is more stable. In the planar polar solution, two +1/2 point defects form at the opposite site of the disk. For low temperature and small $\varepsilon$, the escape radial solution can be obtained. It is a non-singular solution in which $\tens{Q}$ is uniaxial everywhere with $s$ being constant and $\bfmath{n}$ being a harmonic map for the given boundary condition. A phase diagram for the three configurations for $k=2$ is shown in Fig.~\ref{fig_2d_phasek2}. For $k=-2$ there are also three solutions as shown in Fig.~\ref{fig_2d2} (d-f).

\begin{figure}
\centering
  \includegraphics[width=.5\textwidth]{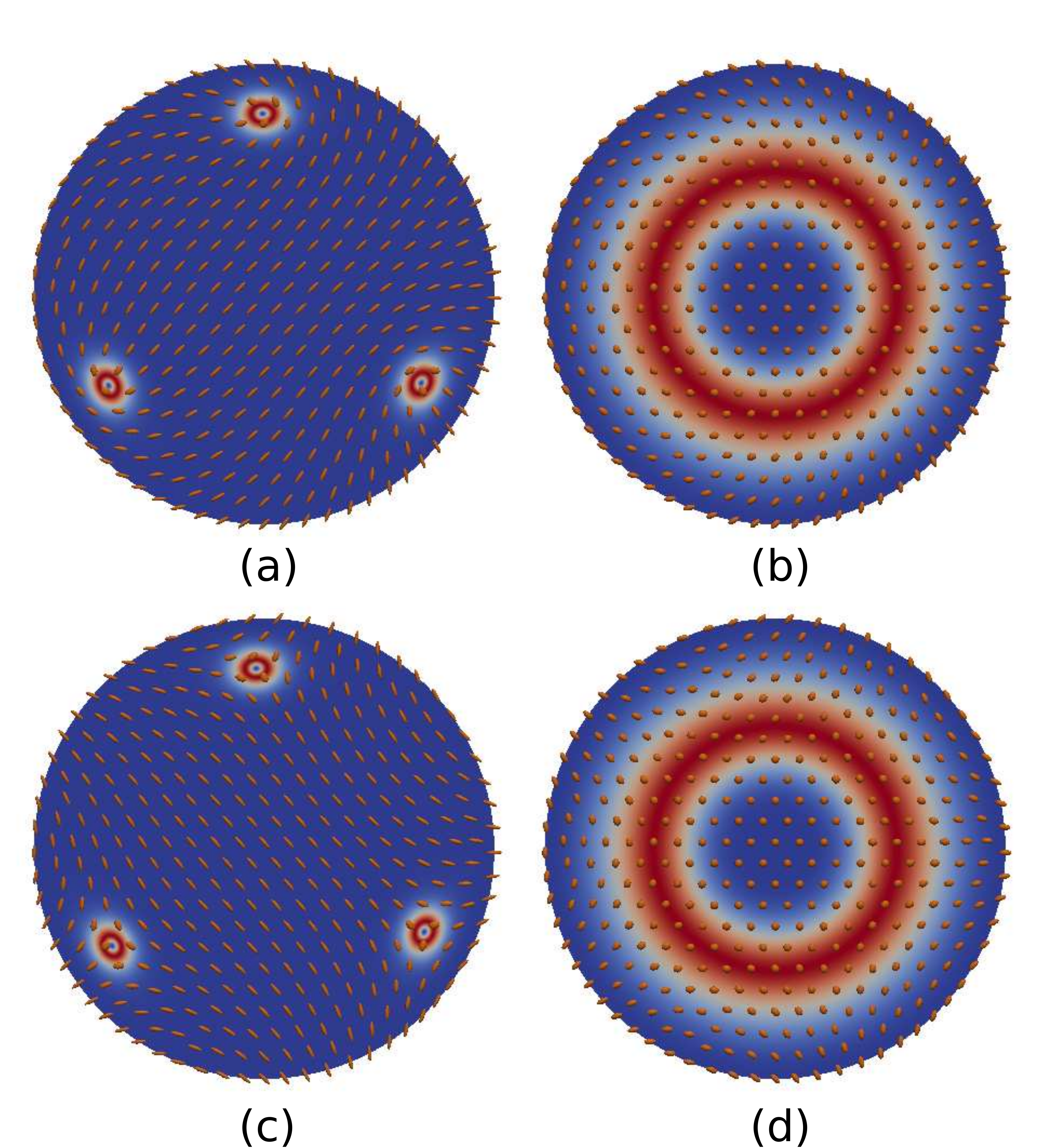}
 \caption{Solutions for $k=3$ (a, b) and $k=-3$ (c, d). $\beta$ is shown in color with red corresponds to biaxial and blue uniaxial. Ellipsoids represent the $\tens{Q}$-tensor. Parameters used are (a, c) $t=-1$, $\varepsilon=0.2$. (b, d) $t=1$, $\varepsilon=0.5$.}
 \label{fig_2d3}
\end{figure}

For $k=\pm 3$, the solutions for different parameters are shown in Fig.~\ref{fig_2d3}. Similar with $k=\pm 2$, at high temperature and large $\varepsilon$, there exist a solution in which the eigenvalues of $\tens{Q}$ are radial symmetry (b, d). It has one point defect with topological charge +3/2 or -3/2 in the center of the disk. At low temperature this point defect will split into three +1/2 or -1/2 point defects (a, c). Unlike the $k=\pm 2$ case, there is no non-singular solution because smooth harmonic map only exists for even $k$ but not for odd $k$.

\begin{figure}
\centering
  \includegraphics[width=.8\textwidth]{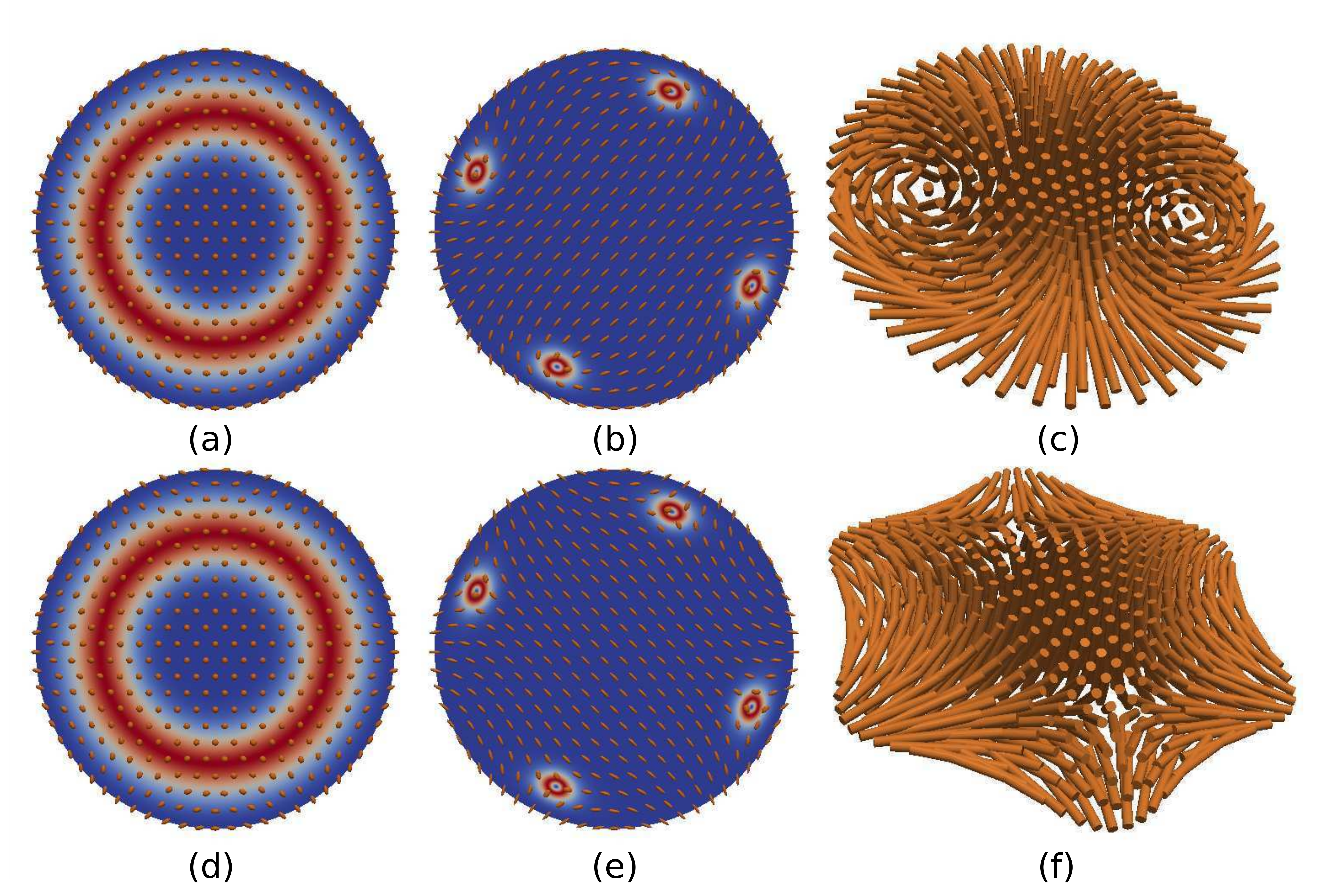}
 \caption{Solutions for $k=4$ (a-c) and $k=-4$ (d-f). $\beta$ is shown in color in (a, b, d, e) with red corresponds to biaxial and blue uniaxial. Ellipsoids represent the $\tens{Q}$-tensor. Parameters used are  (a, d) $t=1$, $\varepsilon=0.5$. (b, e) $t=-1$, $\varepsilon=0.2$. (c, f) $t=-6$, $\varepsilon = 0.1$.}
 \label{fig_2d4}
\end{figure}

The case for $k=\pm 4$ is shown in Fig.~\ref{fig_2d4}. Again, the $\pm 2$ point defect at the center will split into 4 $\pm 1/2$ defect points for low temperature and small $\varepsilon$. There is also a non-singular solutions (c and f) because $k$ is even in this case.

\begin{figure}
\centering
  \includegraphics[width=.8\textwidth]{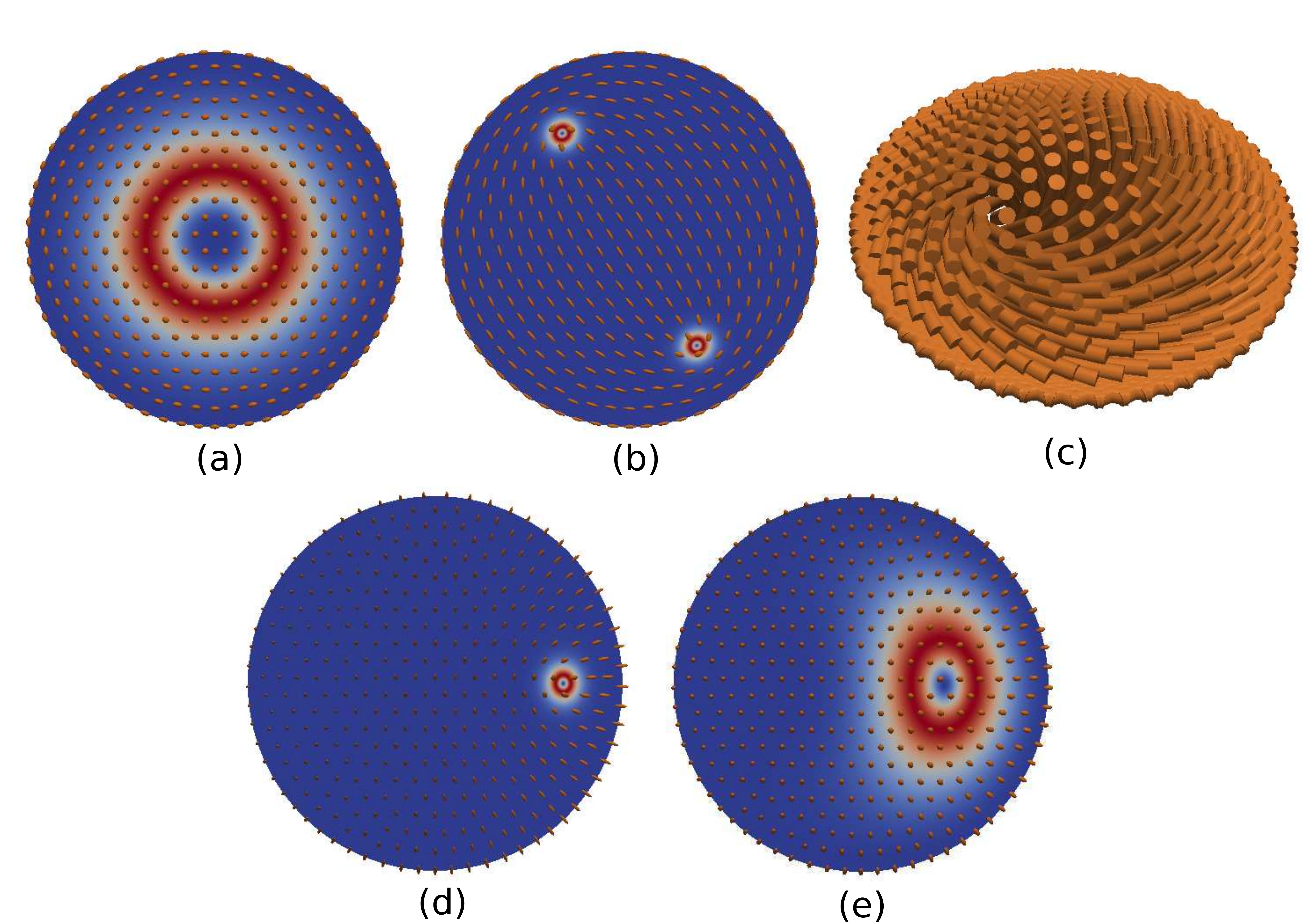}
 \caption{Solutions for tangent (a, b, c) and Mobius (d, e) anchoring condition. (a) Semi-radial solution. $t=1$, $\varepsilon=1$. (a) Two 1/2 point defects in tangent anchoring condition. $t=-1$, $\varepsilon=0.17$. (c) Uniaxial solution. $t=-7$, $\varepsilon=0.033$. (d) Mobius anchoring condition. $t=-1$, $\varepsilon=0.17$. (e) Mobius anchoring condition. $t=1$, $\varepsilon=1$.}
 \label{fig_2dtm}
\end{figure}

We also considered two other boundary conditions. One is the tangent anchoring condition in which $\bfmath{n}$ lies in the tangent direction at the boundary. The other is a Mobius-like anchoring condition in which $\bfmath{n}$ rotates $\pi$ counter-clockwisely in the moving plane traveling alone the boundary circle perpendicular to it (and hence the trajectory of the unit-vector $\bfmath{n}$ forms a Mobius stripe). As shown in Fig.~\ref{fig_2dtm} (a), under the tangent anchoring condition, there is also a radial-symmetrical solution for high temperature and large $\varepsilon$. Again, for low temperature and small $\varepsilon$ two +1/2 point defects will appear (Fig.~\ref{fig_2dtm} (b)). For low temperature and very small $\varepsilon$, there is also a non-singular harmonic map solution (Fig.~\ref{fig_2dtm} (c)). The solutions of the Mobius anchoring condition has a biaxial region located away from the center of the disk (Fig.~\ref{fig_2dtm} (d) and (e)). The profile looks like a +1/2 point defects but the eigenvectors of $\tens{Q}$ are distorted near the defect and no longer perpendicular to or lie in the disk as in other 1/2 point defects.  

To summarize the results in the section, we point out here that, if $\bfmath{n}$ was kept in the $xy$-plane at the boundary (as in the boundary condition Eq.~\eqref{eq_bc2d} and the tangent anchoring condition), the solutions of the LdG model seem to be quite predictable: For large $t$ and $\varepsilon$, there is a \emph{semi-radial} solution in which all the eigenvalues are radial symmetric while the eigenvectors are constant along the $r$-direction up to the boundary. In these cases, $\tens{Q}$ is uniaxial at the boundary (being prolate) as well at the center (being oblate), and there is a connecting biaxial region in between. At the center, there is a defect whose topological charge is determined by the boundary constraint. As $t$ and $\varepsilon$ decrease, the semi-radial solution become unstable and the defect point in the center will quantize to $\pm 1/2$ defect points. The number of $\pm 1/2$ defects is determined by the conservation of the total topological charge. When $k$ is even in Eq.~\eqref{eq_bc2d} and for the tangent anchoring condition, $\bfmath{n}$ can be extended smoothly from the boundary to the entire domain. If this is the case the system admits a  non-singular harmonic map solution, a phenomena referred as ``escape in the third dimension'' in~\cite{sonnet1995alignment}. Both the harmonic map and the quantized-$\pm1/2$ solutions are stable for low temperature. In the limit of $t\rightarrow -\infty$, the free-energy of the former (if exists) will be lower. For boundary conditions in which $\bfmath{n}$ does not stay in the $xy$-plane the behavior of the solution is not fully understood and will be studied in future work.

\section{Profile of disclination lines}
\label{sec_III}
Profile of defect gives the local structure of the $\tens{Q}$-tensor field near the defect. Because defects that are homotopically equivalent to each other should have similar local structure, it is useful to study the profile of some representative defects. The radial hedgehog solution is a good represent for point defects. The profile of the radial hedgehog in the LdG model can be obtained analytically, which is the solution of a second-order ordinary differential equation (ODE). This ODE follows from the radial symmetry property of the radial hedgehog solution plus $\tens{Q}$ is everywhere uniaxial~\cite{majumdar2012radial}. 

For disclination lines, the solutions we obtain for the two-dimensional disk make good representatives. For example, the +1 disclination lines we observed in Fig.~\ref{fig_bc1_three} (c) and Fig.~\ref{fig_bc3_three} (a) in the unit ball are locally homotopically equivalent to the semi-radial solution for $k=2$ (here and in the following \emph{locally} means we are only compare one infinitely small segment of the two disclination lines). Also, the +1/2-disclination lines in Fig.~\ref{fig_bc1_three} (b) and Fig.~\ref{fig_bc3_three} (b, c) in the unit ball are locally homotopically equivalent to the semi-radial solution for $k=1$.

In the following we study the profile of the semi-radial solution for $k = \pm 1, \pm 2, \cdots$. Based on the previous numerical results, we make the following observations:
\begin{enumerate}
 \item There is one and only one defect point located at the center of the disk.
 \item The eigenvectors of $\tens{Q}$ does not change along the $r$-direction for fix azimuth angle $\phi$.
 \item The eigenvalues of $\tens{Q}$ are determined by $r$ only.
\end{enumerate}
These features are mostly evident from Fig.~\ref{fig_2d1} (b) and (d) for the case of $k = \pm 1$ and are also true for other $k$. Based on them we can write $\tens{Q}$ as
\begin{equation}
\tens{Q}(r, \phi) = \lambda_1(r) \bfmath{n}_1(\phi) \bfmath{n}_1(\phi) + 
\lambda_2(r) \bfmath{n}_2(\phi) \bfmath{n}_2(\phi) +
\lambda_3(r) \bfmath{n}_3(\phi) \bfmath{n}_3(\phi),
\label{eq_2danzats} 
\end{equation}
with $0 \le r \le 1$ and $0 \le \phi < 2\pi$. Here $\lambda_1 \ge \lambda_2 \ge \lambda_3$ are the three eigenvalues of $\tens{Q}$ and $\bfmath{n}_1, \bfmath{n}_2, \bfmath{n}_3$ are the corresponding eigenvectors. The eigenvectors are determined by their values at the boundary.

\begin{figure}
  \centering
 \includegraphics[width=.8\textwidth]{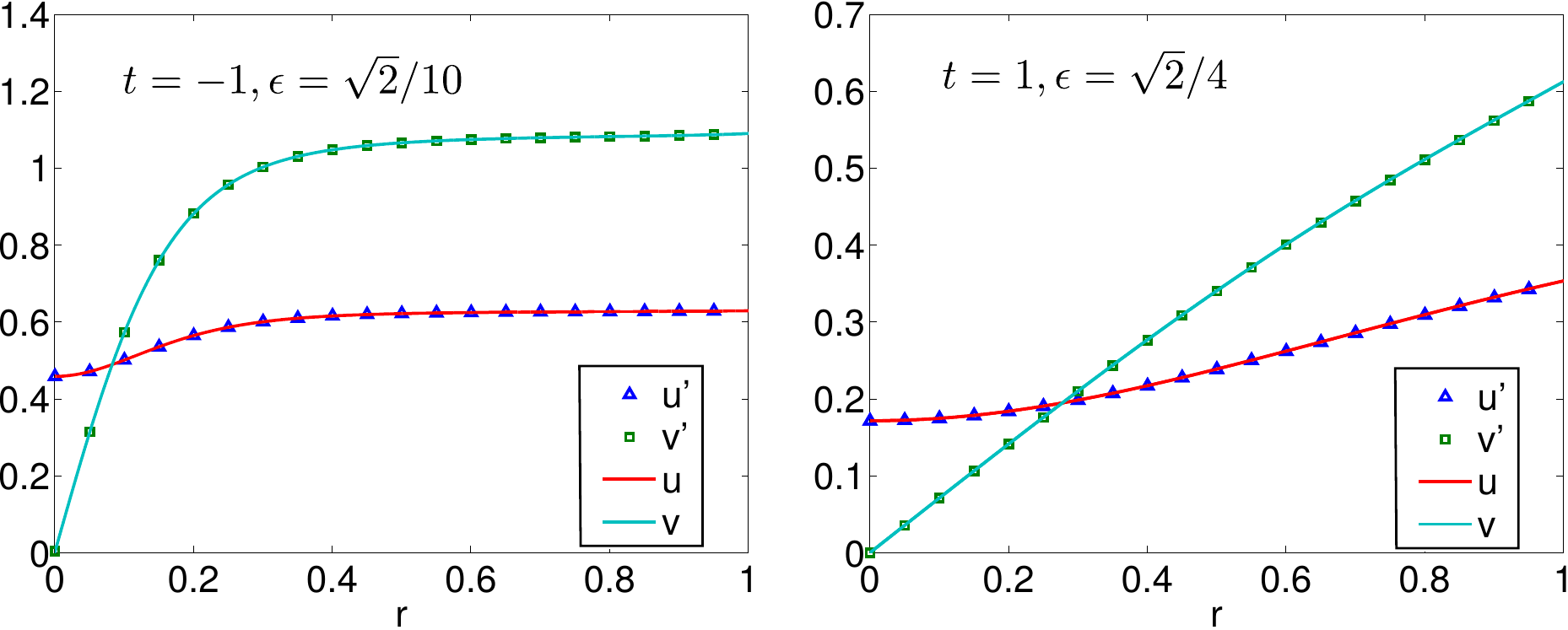}
 \caption{Comparison of the ODEs (Eqs.~\eqref{eq_ode1} and~\eqref{eq_ode2}) and numerical solution. Two sets of parameters are used.}
 \label{fig_testode}
\end{figure}

After change of variables by letting $u = \frac{\sqrt{3}}{2}(\lambda_1 + \lambda_2)$ and $v = \frac{\lambda_1 - \lambda_2}{2}$, $\tens{Q}$ becomes
$$
\tens{Q}(r, \phi) = \left( \begin{array}{ccc}
\frac{\sqrt{3}}{3}u(r) + v(r) \cos(\phi) & v(r) \sin (\phi) & 0 \\
v(r) \sin (\phi) & \frac{\sqrt{3}}{3}u(r) - v(r) \cos(\phi) & 0 \\
0 & 0 & -\frac{2\sqrt{3}}{3}u(r) \end{array} \right).
$$
Substitute this $\tens{Q}$ into the LdG energy function Eq.~\eqref{eq_totalenergy} gives
\begin{eqnarray*}
&& F(\tens{Q}) = F(u(r), v(r)) = \\
&& 2\pi \int_{0}^{1} \left[ t(u^2 + v^2) + 2(u^4 + v^4 + \sqrt{2}u^3 + 2u^2v^2 - 3\sqrt{2}uv^2)
+ 2\varepsilon^2\left( u_r^2 + v_r^2 + \frac{k^2 v^2}{r^2}\right)\right] r dr.
\end{eqnarray*}
The corresponding Euler-Lagrange equation is,
\begin{eqnarray}
 2\varepsilon^2 (u_{rr} + \frac{1}{r}u_{r}) & = & 
tu + 4u^3 + 3\sqrt{2}u^2 + 4uv^2 - 3\sqrt{2}v^2, \label{eq_ode1} \\
2\varepsilon^2 (v_{rr} + \frac{1}{r}v_{r}) & = & 
tv + 4v^3 + 4u^2v - 6\sqrt{2}uv + \frac{2 k^2 \varepsilon^2}{r^2}v. \label{eq_ode2}
\end{eqnarray}
with the boundary condition $u(1) = \sqrt{3}s^+/6$, $v(1) = \frac12 s^+$, $v(0) = 0$, $u'(0)=0$ (The first two conditions comes from Eq.~\eqref{eq_bc2d} at the boundary $r = 1$. The other two conditions is needed for the ODEs to be well-defined at $r=0$). To verify the above results, we solve Eqs.~\eqref{eq_ode1} and~\eqref{eq_ode2} with $k=1$ for different $\varepsilon$ and $t$ and compare the solutions to the numerical results in Fig.~\ref{fig_testode}. The numerical results match the ODEs perfectly.

\begin{remark}
Condition in Eq.~\eqref{eq_2danzats} was proposed in~\cite{fratta2014profiles} for the purpose of obtaining a special solution of the LdG model. Our result is similar to theirs (the meaning of the variables are different), but our motivation is different: we obtain Eq.~\eqref{eq_2danzats} based on observations of numerical results.
\end{remark}

To obtain the profile of disclination lines, we rescale the above ODEs by defining $\tilde{r} = r/(\sqrt{2}\varepsilon)$, $\tilde{u}(\tilde{r}) = u(\sqrt{2}\varepsilon\tilde{r})$ and $\tilde{v}(\tilde{r}) = v(\sqrt{2}\varepsilon\tilde{r})$, let $\varepsilon \to 0$, and drop all the tildes for convenience to give
\begin{eqnarray*}
u_{rr} + \frac{1}{r}u_{r} & = & 
tu + 4u^3 + 3\sqrt{2}u^2 + 4uv^2 - 3\sqrt{2}v^2, \\
v_{rr} + \frac{1}{r}v_{r} & = & 
tv + 4v^3 + 4u^2v - 6\sqrt{2}uv + \frac{k^2}{r^2}v.
\end{eqnarray*}
with the boundary condition $u(+\infty) = \frac{\sqrt{3}}{6}s^+$, $v(+\infty) = \frac{1}{2}s^+$. $v(0) = 0$, $u'(0) = 0$. This solution of the above ODEs gives us a profile that is homotopically equivalent to the $k/2$-disclination lines.

\section{Relation between the LdG and OF model}
\label{sec_IV}
We make a brief discussion about the relation between the LdG tensor model and the Oseen-Frank vector model. In particular, we consider a modified LdG energy functional, given by
\begin{equation}
 F_{mLdG}(\tens{Q}) = \int_{\Omega} \frac{2(f_b(\tens{Q}) - f(s^+))}{\varepsilon^2} + \tens{Q}_{ij,k} \tens{Q}_{ij,k} d\bfmath{x}, 
 \ \   \tens{Q} \in W^{1,2}(\Omega, \mycal{S}_0).
\label{eq_IV1}
 \end{equation}
Here the subtraction of a constant $f(s^+)$ from the bulk energy $f_b(\tens{Q})$ make sure the first term on the right being positive. Given $\varepsilon$, the energy minimizer of Eq.~\eqref{eq_IV1} is denoted by $\tens{Q}^{(\varepsilon)}$. $\tens{Q}^{(\varepsilon)}$ is a solution of the following Euler-Lagrange equation
\begin{equation}
\varepsilon^2 \Delta \tens{Q} = A \tens{Q} - B (\tens{Q}^2 - \frac{\tens{I}}{3}\mathrm{tr}\tens{Q}^2) + C(\mathrm{tr} \tens{Q}^2) \tens{Q}.
 \label{eq_IV3}
\end{equation}
We are interested in the limit of $\tens{Q}^{(\varepsilon)}$ as $\varepsilon \rightarrow 0$.

We define the \emph{limiting harmonic map}
\begin{equation}
\tens{Q}^0 = s^+ \left(\bfmath{n}\bfmath{n} - \frac{\tens{I}}{3} \right),
\label{eq_IV5}
\end{equation}
where $\bfmath{n} \in W^{1,2}(\Omega, \mathbb{R}P^2)$ is the minimizer of the Oseen-Frank free-energy,
\begin{equation}
 F_{OF}(\bfmath{n}) = \int_{\Omega} |\nabla \bfmath{n}|^2 d\bfmath{x}.
\label{eq_IV2}
 \end{equation}
The admissible space of $\bfmath{n}$, $\mathbb{R}P^2 := \mathbb{S}^2/\sim{}$, is the quotient of $\mathbb{S}^2$ with respect to the equivalence relation $\bfmath{n} \sim \bfmath{m}$ if and only if $\bfmath{n} = \pm \bfmath{m}$~\cite{do1992riemannian}.
$\bfmath{n}$ satisfies the following Euler-Lagrange equation,
\begin{equation}
 \Delta \bfmath{n} = - |\nabla \bfmath{n}|^2 \bfmath{n}.
 \label{eq_IV4}
\end{equation}

In the LdG theory, solution of Eq.~\eqref{eq_IV3} may contain point defects and disclination lines. For small but finite $\varepsilon$, and when the temperature is low, disclination lines tend to be more stable than point defects. Moreover, among disclination lines with different topological charge, the one with smaller topological charge tend to be more stable. Since $1/2$ is the smallest topological charge in $\mathbb{R}P^2$, we suspect that disclination lines with topological charge $\pm 1/2$ are generic structure in global energy minimizer $\tens{Q}^{(\varepsilon)}$ of LdG for $\varepsilon > 0$ for arbitrary boundary conditions.

In the case of 3-ball, it is shown in [21] that, under Dirichlet boundary condition, there exists a sequence of global minimizer $\tens{Q}^{(\varepsilon)}$ of the LdG such that $\tens{Q}^{(\varepsilon)} \rightarrow \tens{Q}^{0}$ as $\varepsilon \rightarrow 0$ strongly in the Sobolev space $W^{1,2}(\Omega, \mathbb{R}P^2)$, where $\tens{Q}^{0}$ is the limiting harmonic map defined in Eq.~\eqref{eq_IV5}. In particular, as we mentioned earlier, the disclination ring will shrink to the radial hedgehog as $\varepsilon\rightarrow 0$, which is achieved by rescaling the size of an infinitely large ball to a unit ball (Fig.~\ref{fig_bc1_rr}). Additionally, for planar condition the two 1/2-defects on the surface will also shrink to zero as $\varepsilon \rightarrow 0$ (Fig.~\ref{fig_rt} (a)). We believe this limiting process is also true for arbitrary boundary conditions.

\begin{figure}
\centering
 \includegraphics[width=0.5\textwidth]{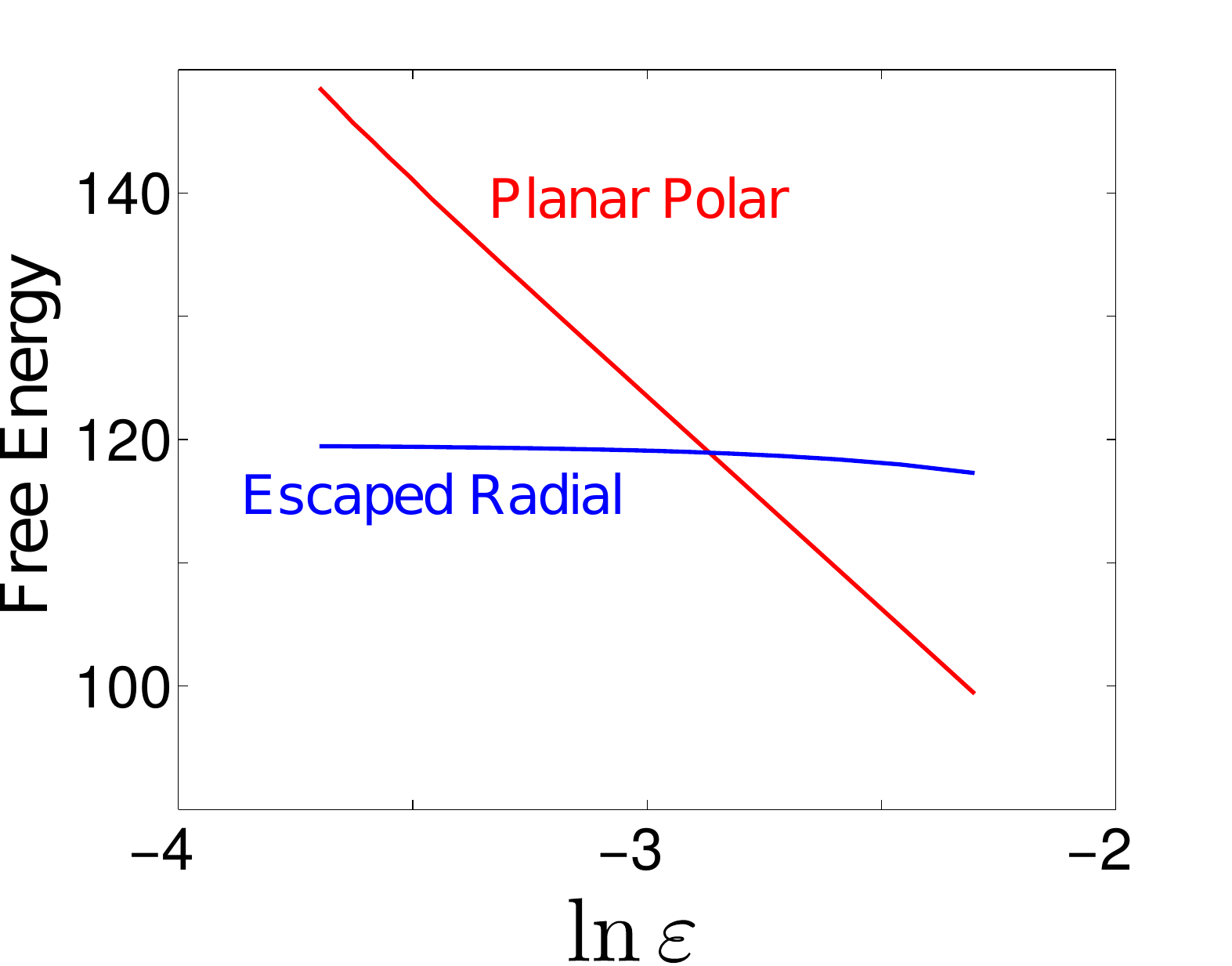}
 \caption{Total free-energy of the planar polar (red) and escaped radial (blue) solution for the two-dimensional disk ($k=2$).}
 \label{fig_2dradial_energy}
\end{figure}

In the case of 2-disk, for sufficiently low temperature, the Euler-Lagrange equation of the LdG functional Eq.~\eqref{eq_IV3} admits a solution $\tens{Q}^{(\varepsilon)}_l$ that contains quantized $\pm1/2$-disclination lines.  $\tens{Q}^{(\varepsilon)}_l$ is meta-stable in the LdG model for any given $t$ and $\varepsilon$, but not a minimizer in the Oseen-Frank model because it contains disclination lines. In some cases, such as when $k$ is even in Eq.~\eqref{eq_bc2d}, $\tens{Q}^{(\varepsilon)}_l$ may ``escape" to the harmonic map solution $Q^{0}$, which is the global minimizer of the LdG free-energy for sufficiently small $t$ and $\varepsilon$ (see, for example, Fig.~\ref{fig_2dradial_energy}). In some other cases, such as when $k$ is odd in Eq.~\eqref{eq_bc2d}, an harmonic map $\tens{Q}^0$ does not exist. If this is the cases then $\tens{Q}^{(\varepsilon)}_l$ will not approach to $\tens{Q}^0$ as $\varepsilon \rightarrow 0$ because in the 2-disk $\tens{Q}$ is assumed to be invariant along the $z$-axis. When this constraint is removed in full three-dimensional cylinder, we expect these infinitely long vertical disclination lines to break up and shrink in size as $\varepsilon\rightarrow 0$. In this sense, the above limit from the LdG model to the Oseen-Frank model still holds.

\section{Discussion and Conclusion}
\label{sec_V}
In this work, we investigate defect pattern in the LdG model in a three-dimensional ball and  two-dimensional disk subject to different boundary conditions. We classify defects into five categories according to their patterns. Among them, only the radial hedgehog solution contains point defect while all other four cases contain disclination lines. A common feature shared by disclination lines is that they are always accompanied by biaxial region. The profile of disclination lines highlight the drastic difference between the tensor model and vector model, and are the focus of this work.

We try to understand the properties of disclination lines using both numerical and analytical approaches. Our numerical results provide detailed configuration of defect under different boundary conditions. Based on observations made from these results, we obtain the profiles of disclination lines analytically in two-dimensional disk. The profiles are important for us to understand the local structure of nematic LCs near defects. 

To summarize the key properties of defect pattern within the LdG theory, four conjectures are proposed in the following.

\begin{conjecture}
Disclination lines are a more generic way for energy concentration than point defects.
\end{conjecture}
Point defects and disclination lines are two forms of defects in the LdG model that are different in topology and local profile. One can also think them as two ways of energy concentration (see Fig.~\ref{fig_bc1_energy}): To minimize the total energy, a large portion of $\Omega$ is kept in the lowest energy state, while most of the excessive energy is concentrated in the vicinity of defects. It appears that forming a 1/2-disclination line is the most efficient way of energy concentration in order to reduce the total energy cost.

With the idea of energy concentration in mind we can understand the previous numerical results in a more systematic way. For the 3-ball under radial anchoring condition, the excessive free-energy is concentrated near the center of the ball. At the temperature decreases, the way of energy concentration will switch from point defect to disclination ring, causing symmetry breaking. For the planar anchoring condition, energy is concentrated near a thin boundary layer while the bulk body of the ball is in the lowest energy state. Within the boundary layer, energy will further distributed to form disclination lines.

For the 2-disk cases, energy concentration explains why disclination lines with high topological charge will quantize to $\pm 1/2$-disclination lines at low temperature and small $\varepsilon$. However, these disclination lines are obtained under the assumption that $\tens{Q}$ is invariant along the $z$-axis. For a infinitely long cylinder of nematic LCs, the energy cost in maintaining these disclination lines is also infinite. As a result, if the boundary condition allows, the solution will escape in the third dimension to a non-singular solution.

\begin{conjecture}
The local configuration of defects can be described by their corresponding profiles.
\end{conjecture}
We obtain the profiles of $k$-disclination lines as given by Eq.~\eqref{eq_ode1} and~\eqref{eq_ode2}. Other disclination lines with the same topological charge in the LdG model will be homotopically equivalent to them and hence their local $\tens{Q}$-tensor field can be described by these profiles.
 
\begin{conjecture}
Among all the disclination lines, the $\pm 1/2$-disclination line is the most stable.
\label{c3}
\end{conjecture}
In both the three- and two-dimensional results, the $\pm k/2$-disclination lines for $k>1$ can only exists for relatively high temperature and large $\varepsilon$. For low $t$ and small $\varepsilon$, they will quantize to give $k$ $\pm 1/2$-disclination line. In a special setting, it can be proven for the two-dimensional disk that the $\pm 1/2$-disclination lines is the most stable structure in the LdG theory~\cite{bauman2012analysis}. We believe this statement is generally true for all tensor models of LCs.

\begin{conjecture}
For a point at disclination line, $\tens{Q}$ is always uniaxial, with $s<0$ (oblate) and $\bfmath{n}$ pointing in the axial direction.
\label{c4}
\end{conjecture}
There might be exceptions to this conjecture under extreme conditions. For example, in the Mobius anchoring condition, conjecture~\ref{c4} seems to be violated in the two-dimensional disk. However, whether this conjecture is true or not for the Mobius anchoring condition in the physically more realistic three-dimensional cylinder is not clear to us. Together, conjectures \ref{c3} and \ref{c4} characterize the geometry properties of disclination lines. 

Overall, the above conjectures give an integrated description of defect pattern, including the global position and local profile. Although they are based on results obtained within the LdG model, we believe they are qualitatively true for other tensor models of LCs. These conjectures open a new perspective in defect pattern of LCs and pose interesting mathematical problems for future research. Both numerical and analytical approaches are needed in order to fully understand these problems.
 
\section*{Acknowledgment}
PZ is supported by the NSFC (National Science Foundation of China) under Grant 21274005. YH is supported by the NSFC under Grant No. 11301294.

\newpage
\noindent\textbf{\LARGE{Appendix}} 
\appendix
\setcounter{figure}{0}
\renewcommand{\thetable}{S\arabic{table}}  
\renewcommand{\thefigure}{S\arabic{figure}}

\section{Expanding $\tens{Q}$ in Zernike polynomials}
For the three-dimensional ball, we expand each element of the $\tens{Q}(r,\theta,\phi)$ using Zernike polynomials,
\begin{equation}
q_i(r,\theta,\phi) = \sum\limits_{m = 1-M}^{M-1}\sum\limits_{l=|m|}^{L-1}\sum\limits_{n=l}^{N-1}A^{(i)}_{nlm}Z_{nlm}(r,\theta,\phi).
\label{eq_zernike}
\end{equation}
where $N \geq L \geq M \geq 0$,
\begin{equation*}
Z_{nlm}(r,\theta,\phi) = R_{n}^{(l)}(r)Y_{lm}(\theta,\phi), 
\end{equation*}
\begin{equation*}
R_{n}^{(l)}(r) = \left\{
\begin{array}{ll}
\sum\limits_{s=0}^{(n-l)/2} N_{nls} r^{n-2s}, & \frac{n-l}{2} \geq 0, \frac{n-l}{2} \in Z \\
0 & others
\end{array}
\right.
.
\end{equation*}
\begin{equation*}
N_{nls} = (-1)^s\sqrt{2n+3}\prod\limits_{i=1}^{n-l}(n+l-2s+1+i)\prod\limits_{i=1}^{l}(\frac{n-l}{2}-s+i)\frac{2^{l-n}}{s!(n-s)!}.
\end{equation*}
$Y_{lm}(\theta,\phi) = P_l^{|m|}(\cos\theta)X_m(\phi)$ are the spherical harmonic functions, 
\begin{equation*}
X_{m}(\phi) = \left\{
\begin{array}{ll}
\cos m\phi, & m \geq 0\\
\sin |m|\phi, & m < 0
\end{array}
\right.
.
\end{equation*}
$P_l^{m}(x)$~$(m \geq 0)$ are the normalized associated Legendre polynomials. $Z_{nlm}$ have the properties:
\begin{equation*}
\int_0^1 \int_0^{2\pi} \int_0^{\pi} Z_{nlm}Z_{n'l'm'}r^2\sin\theta d\theta d\phi dr = \delta_{nn'}\delta_{ll'}\delta_{mm'}.
\end{equation*}
\begin{equation*}
\int_0^1 \int_0^{2\pi} \int_0^{\pi} \nabla Z_{nlm}\cdot\nabla Z_{n'l'm'}r^2\sin\theta d\theta d\phi dr = \delta_{ll'}\delta_{mm'}K_{nn'l}.
\end{equation*}
where
\begin{equation*}
K_{nn'l} = \int_0^1 \frac{d R_{n}^{(l)}}{dr}\frac{d R_{n'}^{(l)}}{dr}r^2 dr + l(l+1)\int_0^1 R_{n}^{(l)}(r)R_{n'}^{(l)}(r)dr.
\end{equation*}

For the two-dimensional disk the procedure is similar. We expand each element of $Q(r,\phi)$ using 2D Zernike polynomials,
\begin{equation}
q_i(r,\phi) = \sum\limits_{m = 1-M}^{M-1}\sum\limits_{n=|m|}^{N-1}A^{(i)}_{nm}Z_{nm}(r,\phi).
\label{eq_zernike2d}
\end{equation}
where 
\begin{equation*}
Z_{nm}(r,\phi) = R_{n}^{(|m|)}(r)X_{m}(\phi), 
\end{equation*}
\begin{equation*}
R_{n}^{(m)}(r) = \left\{
\begin{array}{ll}
\sum\limits_{s=0}^{(n-m)/2} \tilde{N}_{nms} r^{n-2s}, & \frac{n-m}{2} \geq 0, \frac{n-m}{2} \in Z \\
0 & others
\end{array}
\right.
.
\end{equation*}
\begin{equation*}
\tilde{N}_{nms} = (-1)^s\sqrt{2n+2}\frac{(n-s)!}{s!(\frac{n+m}{2}-s)!(\frac{n-m}{2}-s)!}.
\end{equation*}
\begin{equation*}
X_{m}(\phi) = \left\{
\begin{array}{lll}
\frac{1}{\pi}\cos m\phi, & m > 0\\
\frac{1}{2\pi}, & m = 0\\
\frac{1}{\pi}\sin |m|\phi, & m < 0
\end{array}
\right.
.
\end{equation*}

\section{Algorithm}
After expanding $\tens{Q}$ in Zernike polynomials, we need to determine the coefficients $A_{nlm}^{(i)}$ in Eq.~\eqref{eq_zernike}. $\tens{Q}$ is a function of $[L/2 \times (N-L/2+1) \times (2M-1)-M/2 \times (M/2-1)-(3N-M+2)\times M\times (M-1)/6]\times 5$ variables (5 is the number of free variables in a three-by-three traceless symmetrical matrix). So does the total free-energy $F$. Given $A_{nlm}^{(i)}$, the integration of $\nabla \tens{Q}$ term can be computed analytically using $A_{nlm}^{(i)}$ from the orthogonal relation of the Zernike polynomials. For the bulk energy, numerical integration is used. In particular, we use Gaussian integral in $r$ and $\theta$ and fast Fourier transform in $\phi$. The calculation of the gradient of $F$ with respect to $A_{nlm}^{(i)}$ is similarly. The gradient information allows us to use optimization methods such as BFGS~\cite{avriel2012nonlinear} to find $A_{nlm}^{(i)}$ that minimize the energy $F$.

\begin{figure}
\centering
\includegraphics[width=.5\textwidth]{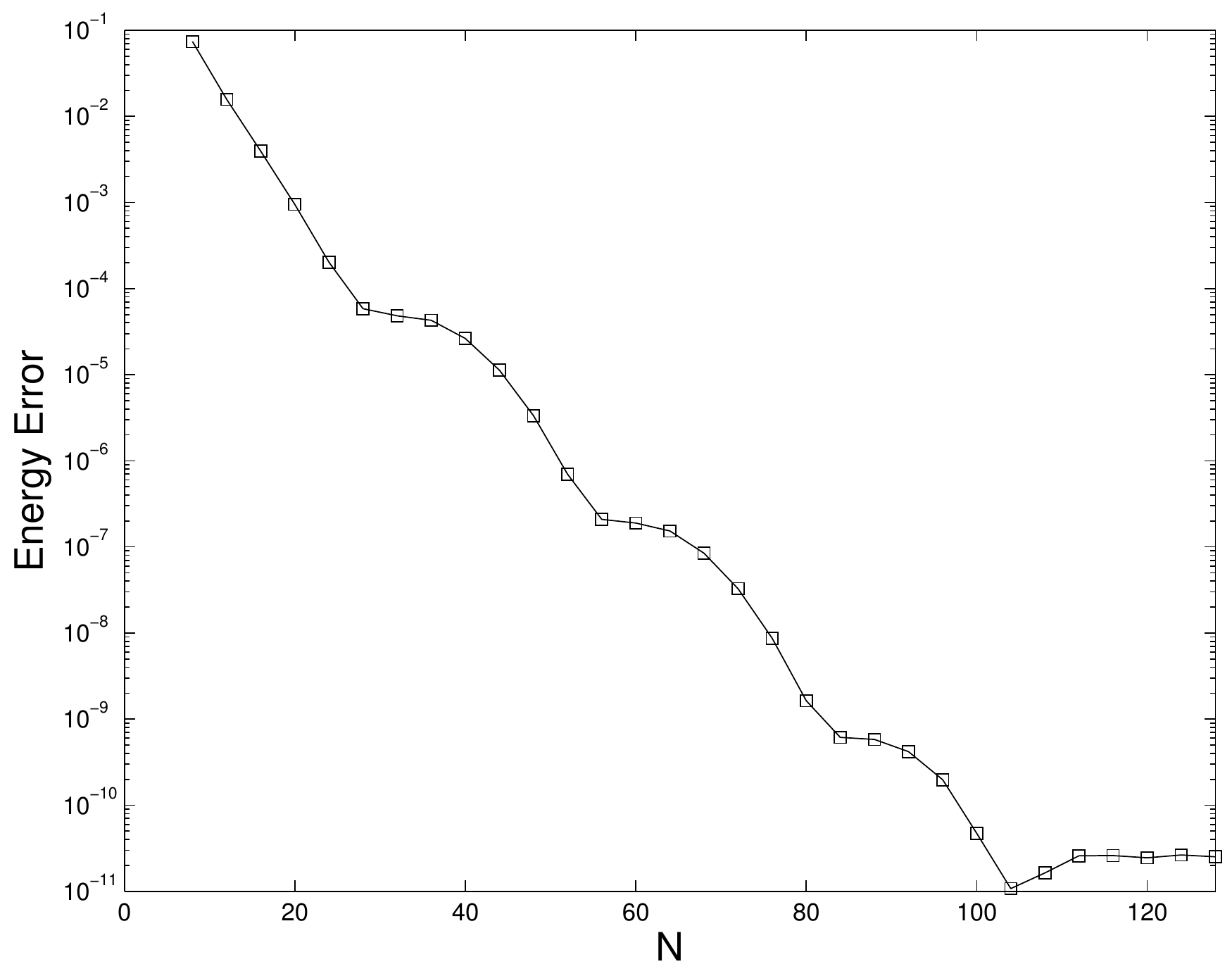}
\caption{Relative error in free-energy for the radial hedgehog solution. $(F_{numer} - F_{exact})/F_{exact}$.}
\label{fig_error}
\end{figure}

The choice of $N, L, M$ are rather arbitrary and can be adjusted to get the best performance. What we did is to start with some moderate $N, L, M$ and gradually increase some or all of them until the numerical solution converge, i.~e., no significant change in the value of free-energy. To validate the algorithm, we compare our numerical results to the radial hedgehog solution. The later can be obtained analytically by assuming radial symmetry~\cite{majumdar2012radial}. As we increase the number of basis in the Zernike polynomials using $N = 4k, L = 16, M = 4$, the numerical error in the total free-energy decrease to as low as $10^{-10}$ (Fig.~\ref{fig_error}).

3D figures in this work are produced using Paraview (http://www.paraview.org/).

\bibliographystyle{siam}
\bibliography{ref}

\end{document}